\documentclass[sigconf]{acmart}
\usepackage{subcaption} 
\usepackage{multirow}
\usepackage{hyperref}
\usepackage{tablefootnote}
\usepackage{cutwin}

\usepackage{amsmath}
\usepackage{wasysym}
\usepackage{xcolor,colortbl}
\usepackage{soul}
\usepackage{enumitem} 

\usepackage{tikz}

\AtBeginDocument{%
  }

\setcopyright{acmlicensed}
\copyrightyear{2018}
\acmYear{2018}
\acmDOI{XXXXXXX.XXXXXXX}
\acmConference[SIGMOD 26]{ACM SIGMOD Conference on Management of Data}{June 03--05,
  2026}{Bangalore, India}
\acmISBN{978-1-4503-XXXX-X/18/06}

\newcommand{\reviewerone}[1]{{\color{black} #1}}
\newcommand{\reviewertwo}[1]{{\color{black} #1}}
\newcommand{\reviewerthree}[1]{{\color{black} #1}}

\begin{document}

\title{Attribute Filtering in Approximate Nearest Neighbor Search: An~In-depth Experimental Study}

\author{Mocheng Li}
\affiliation{%
  \institution{The Chinese University of Hong Kong, Shenzhen}
  \city{Shenzhen}
  \country{China}
}
\email{mochengli1@link.cuhk.edu.cn}

\author{Xiao Yan}
\affiliation{%
  \institution{Institute for Math \& AI, Wuhan, Wuhan University}
  \city{Wuhan}
  \country{China}}
\email{yanxiaosunny@gmail.com}

\author{Baotong Lu}
\affiliation{%
  \institution{Microsoft Research}
  \city{Beijing}
  \country{China}
}
\email{baotonglu@microsoft.com}

\author{Yue Zhang}
\affiliation{%
  \institution{The Chinese University of Hong Kong, Shenzhen}
  \city{Shenzhen}
  \country{China}
}
\email{223040247@link.cuhk.edu.cn}

\author{James Cheng}
\affiliation{%
 \institution{The Chinese University of Hong Kong}
  \city{Hong Kong}
  \country{China}
 }
\email{jcheng@cse.cuhk.edu.hk}

\author{Chenhao Ma}
\affiliation{%
  \institution{The Chinese University of Hong Kong, Shenzhen}
  \city{Shenzhen}
  \country{China}
}
\email{machenhao@cuhk.edu.cn}
\authornote{Corresponding author.}

\renewcommand{\shortauthors}{Mocheng Li, Xiao Yan, Baotong Lu, Yue Zhang, James Cheng and Chenhao Ma.}

\begin{abstract}

With the growing integration of structured and unstructured data, new methods have emerged for performing similarity searches on vectors while honoring structured attribute constraints, i.e., a process known as Filtering Approximate Nearest Neighbor (Filtering ANN) search. Since many of these algorithms have only appeared in recent years and are designed to work with a variety of base indexing methods and filtering strategies, there is a pressing need for a unified analysis that identifies their core techniques and enables meaningful comparisons.

In this work, we present a unified Filtering ANN search interface that encompasses the latest algorithms and evaluate them extensively from multiple perspectives. First, we propose a comprehensive taxonomy of existing Filtering ANN algorithms based on attribute types and filtering strategies. Next, we analyze their key components, i.e., index structures, pruning strategies, and entry point selection, to elucidate design differences and tradeoffs. We then conduct a broad experimental evaluation on 10 algorithms and \reviewertwo{12} methods across 4 datasets (each with up to 10 million items), incorporating both synthetic and real attributes and covering selectivity levels from 0.1\% to 100\%. Finally, an in-depth component analysis reveals the influence of pruning, entry point selection, and edge filtering costs on overall performance. Based on our findings, we summarize the strengths and limitations of each approach, provide practical guidelines for selecting appropriate methods, and suggest promising directions for future research. Our code is available at: \url{https://github.com/lmccccc/FANNBench}.

\end{abstract}



\received{13 March 2025}
\received[revised]{1 July 2025}
\received[accepted]{24 August 2025}

\maketitle

\section{Introduction}

\begin{figure}[t]
    \centering
    \begin{subfigure}[b]{0.9\linewidth}
        \includegraphics[width=\linewidth]{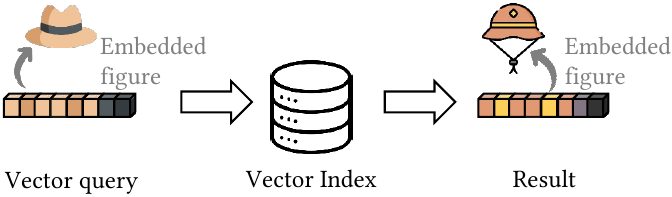}
        \caption{ANN search}
        \label{ANN}
    \end{subfigure}
    \hfill
    \begin{subfigure}[b]{0.9\linewidth}
        \includegraphics[width=\linewidth]{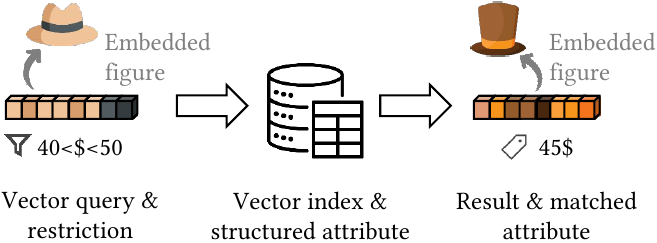}
        \caption{Filtering ANN search}
        \label{FANN}
    \end{subfigure}
    \caption{Example of ANN search and Filtering ANN search.}
    \label{ANN_example}
\end{figure}

The advent of Transformer architectures~\cite{vaswani2017attention} and modern embedding techniques~\cite{radford2021learning} has led to the widespread use of vector representations for unstructured data such as text, video, audio, and images. 
These breakthroughs in representation learning have driven rapid progress in {\bf Approximate Nearest Neighbor (ANN)} search technologies, whose core objective is to efficiently retrieve the top-$k$ vectors that are approximately most similar to a given query vector (Figure \ref{ANN}).
ANN search is integral to vector databases~\cite{wei2020analyticdb, zhang2023vbase, wang2021milvus, pinecone, mohoney2023high,sptag,ngt}, serving as the fundamental system for recommendation systems~\cite{meng2020pmd, okura2017embedding, paterek2007improving}, pattern matching~\cite{cover1967nearest, kosuge2019object}, and retrieval-augmented generation (RAG)~\cite{edge2024local, he2024g, gutierrez2024hipporag}. 

Various indexing methods have been developed for efficient ANN search, including graph-based~\cite{wang2024efficient, jayaram2019diskann}, quantization-based~\cite{jegou2010product, gao2024rabitq}, hashing-based~\cite{datar2004locality, gong2020idec}, and tree-based~\cite{arora2018hd, silpa2008optimised} approaches. 
According to recent academic and industrial studies~\cite{wang2021comprehensive, aumuller2020ann, yang2024revisiting, schafer2024fast}, graph-based and quantization-based methods are most prominent. 
Graph-based algorithms construct neighborhood graphs that offer high query efficiency and recall by traversing from an entry point to successively closer neighbors~\cite{li2019approximate, aurenhammer2013voronoi, toussaint1980relative, paredes2005using, kruskal1956shortest}. 
On the other hand, quantization-based methods compress vectors—via dimensionality reduction~\cite{jaasaari2024lorann} or lower-bit encoding~\cite{niu2023residual, pan2020product}—resulting in low memory usage, good GPU parallelism compatibility, and often integrate inverted file structures for enhanced efficiency.

\subsection{Filtering ANN Search}

Building upon the foundation of ANN search, there is a growing need for hybrid ANN search scenarios that integrate vector similarity search with relational database-style functionalities.
This hybrid search treats vectors as a native data type while enabling SQL-like queries over structured attributes.
For example, one might search for visually similar products while restricting results by price (see Figure~\ref{FANN}) or retrieve similar photos taken by the same individual. 
We refer to this hybrid search task as \textbf{Filtering Approximate Nearest Neighbor (Filtering ANN)} search. 


Various Filtering ANN search methods have been developed through multiple integration strategies between attribute filtering and vector similarity search~\cite{matsui2018reconfigurable, pgvector, qdrant}. 
Moreover, some methods focus on specific attribute types for better performance~\cite{gollapudi2023filtered, zuo2024serf, wang2024efficient}, but this complicates the classification of filtering tasks.
To effectively handle different attribute types, we provide a systematic taxonomy from two perspectives: attribute types and filtering strategies.

\vspace{1em}

\begin{table}
  \caption{Attribute types and corresponding filtering types.}
  \label{fmethod}
  \begin{tabular}{ccc}
    \toprule
     Attribute Type & Filtering Type & Query Predicates \\
    \hline
    Numerical & Range & $l \leq O_i.a \leq u$ \\
    Categorical & Label & $f \in O_i.A$ \\
    Arbitrary & General & Any filtering expression \\
  \bottomrule
\end{tabular}
\end{table}

\textbf{Filtering Attributes.} Current systems or methods predominantly target two attribute types: \textit{Numerical} and \textit{Categorical}, corresponding to \textit{Range Filtering ANN search} and \textit{Label Filtering ANN search}, respectively.

\begin{itemize}[leftmargin=*, nosep]
    \item \textit{Numerical Attributes and Range Filtering.} Numerical attributes, such as price or date, represent continuous values. Range filtering retrieves nearest neighbors whose attribute values fall within a specified interval (e.g., products within a price range or photos taken between two dates). For a range Filtering ANN query, let $O_i$ denote a vector–attribute pair $(v, a)$ with index $i$, where $O_i.a$ represents its numerical attribute. Let $l$ and $u$ denote the query's lower and upper bounds, respectively, as shown in Table~\ref{fmethod}. Several approaches, including SeRF~\cite{zuo2024serf}, iRangeGraph~\cite{xu2024irangegraph}, and UNIFY~\cite{liang2024unify}, have been designed to address this scenario.

    \item \textit{Categorical Attributes and Label Filtering.} Categorical attributes represent discrete labels. For example, YouTube videos often carry tags like ``music'', ``comedy'', or ``news'' that describe their content. Label filtering finds nearest neighbors (i.e., similar videos) that share one or more of these tags. In Table~\ref{fmethod}, $f$ denotes the query’s attribute, while $O_i.A$ represents the categorical attribute set of $O_i$. Filtered DiskANN~\cite{gollapudi2023filtered} is a classic example of this approach.

    \item \textit{Arbitrary Filtering.} Beyond categorical and numerical filters, some systems support more flexible filtering restrictions via a user-defined filtering function. These methods either integrate closely with traditional databases, e.g., VBASE~\cite{zhang2023vbase} and Milvus~\cite{wang2021milvus} or manage arbitrary subset searches while abstracting away filtering details, e.g., Faiss~\cite{douze2024faiss} and ACORN~\cite{patel2024acorn}.

\end{itemize}

\vspace{1em}

\begin{figure}[t]
    \centering
    \includegraphics[width=\linewidth]{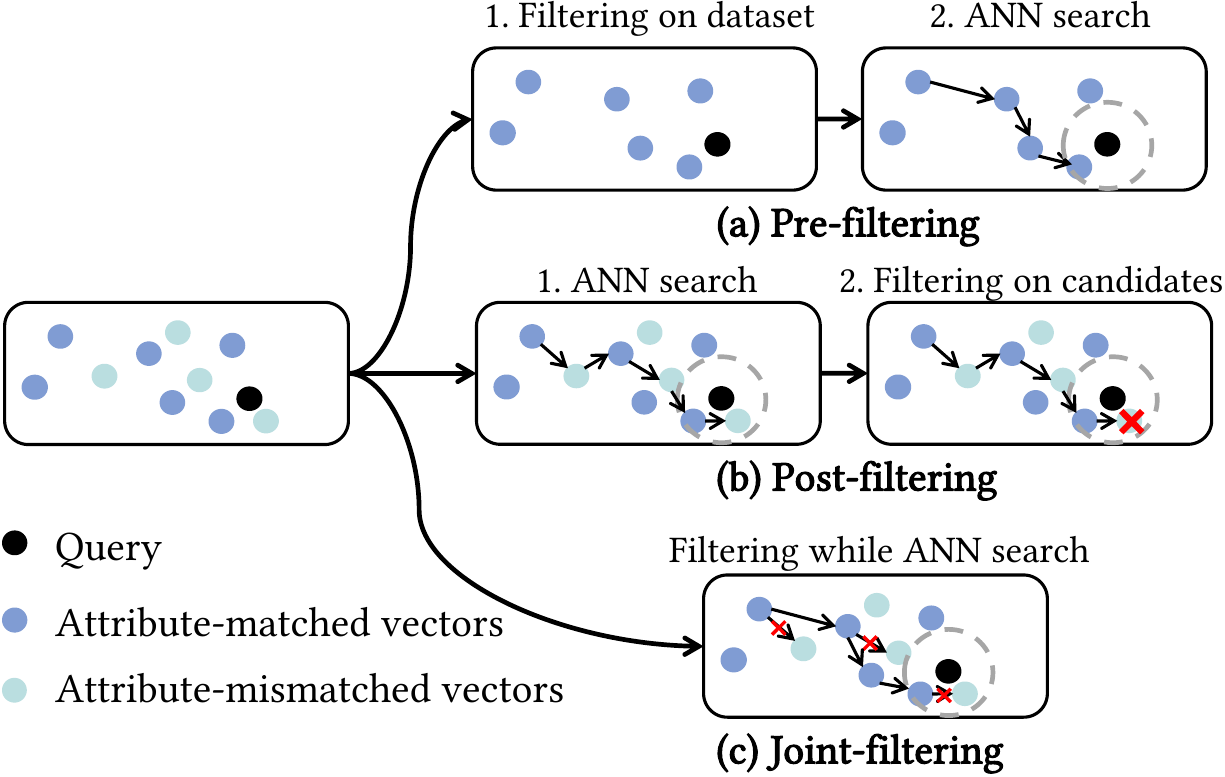}
    \caption{ANN filtering strategies.}
    
    \label{filtering_methods}
\end{figure}

\textbf{Filtering Strategies.} Figure~\ref{filtering_methods} illustrates common approaches to integrating attribute filtering with ANN search. These approaches differ in the order in which filtering is applied relative to the ANN search process, and we classify them into three categories: {\em pre-filtering}, {\em post-filtering}, and {\em joint-filtering}. In this context, {\em selectivity} refers to the fraction of data items that pass a filtering condition: high selectivity means many pass, while low selectivity means few do.

\begin{itemize}[leftmargin=*, nosep]
    \item \textit{Pre-filtering} applies filtering prior to the similarity search to reduce the search space—effective for data-independent indexes such as the Inverted File Index (IVF). However, in graph-based methods, pruning nodes based on attributes may disrupt traversal paths and reduce recall.

    \item \textit{Post-filtering} performs the ANN search first and then removes candidates that do not meet the attribute constraints. This approach is efficient when filter selectivity is high but becomes costly in low-selectivity scenarios, as many invalid candidates need to be discarded after the search.

    \item \textit{Joint-filtering} integrates attribute filtering directly into the ANN search. For example, in a graph-based search, the traversal is restricted to edges connecting to nodes that satisfy the attribute constraints, allowing the algorithm to prune paths early. This dynamic approach is especially effective when filter selectivity is moderate.

\end{itemize}

\subsection{Our Contributions}

Motivated by the rising demand for efficient Filtering ANN algorithms and the diversity of existing filtering strategies, we conduct a comprehensive experimental survey to clarify their strengths, limitations, practical application scenarios, and open problems. Our contributions are summarized as follows:

\textbf{Taxonomy Study \reviewertwo{(Section~\ref{sec:overview})}.}  
While numerous methods have reported impressive performance gains, it remains challenging to identify their core ideas, advantages, and limitations. We systematically review \reviewertwo{12} systems and algorithms, providing a unified classification based on their indexing methods, filtering strategies, and key techniques \reviewertwo{(Figure~\ref{roadmap} and Table~\ref{overviewFANN})}. Our two-axis taxonomy categorizes methods by filtering type (label, range, and arbitrary filtering) and by filtering strategy (pre-filtering, post-filtering, and joint-filtering), offering clear insights into their design trade-offs.

\textbf{Comprehensive Experimental Evaluation \reviewertwo{(Section~\ref{sec:exp:range},~\ref{sec:exp:label} and~\ref{sec:exp:index})}.}  
Given that most algorithms in this area have been published recently and involve numerous hyperparameters, comparisons remain challenging. 
To address this, we develop a unified interface that enables consistent evaluation, and we conduct extensive experiments across 12 methods from 10 representative papers, covering both label and range filtering tasks with query selectivity ranging from 0.1\% to 100\%. Our results indicate that segmented graph indexing excels in range filtering scenarios, while label filtering techniques are often unstable due to suboptimal graph quality.

\textbf{Component Analysis \reviewertwo{(Section~\ref{sec:component}, \ref{sec:exp:prune}, \ref{sec:exp:entry}, and \ref{sec:exp:edgesel})}.}  
We analyze the core components of Filtering ANN algorithms to determine which factors contribute most significantly to overall performance. Our analysis reveals that constructing an index for all possible subsets is both intuitive and effective. Additionally, we identify underexplored components in graph-based methods, such as pruning strategies and entry point selection, that are 
crucial for performance. 
We have several interesting findings: (1) Relative Neighbor Graph (RNG) pruning breaks down at low selectivity; (2) hierarchical multi-layer indexes boost performance only at high selectivity and are otherwise optional; and (3) bypassing hierarchy in these indexes by increasing bottom‑layer entry points consistently improves performance across all selectivity levels.


\textbf{Usage and Development Guidelines \reviewertwo{(Section~\ref{sec:lessons})}.}  
Recognizing that Filtering ANN algorithms are designed for distinct scenarios, we propose practical guidelines (Figure~\ref{guidebook}) to help practitioners select the most suitable method based on specific filtering needs and data characteristics. We also discuss emerging trends and potential future research directions to drive further innovations in this field.

\section{Preliminaries}

\begin{table}[t]
  \caption{Notations in our paper.}
  \label{symbol}
  \begin{tabular}{l|l|l}
    \hline
    \multicolumn{2}{c|}{\textbf{Notations}} & \textbf{Descriptions}  \\
    \hline
    \multicolumn{2}{c|}{$\mathcal{D}=\{O\}$} & Vector dataset with attributes. \\
    \hline
    \multirow{2}{*}{$O$} & $=(v,a)$ & Vector and numerical attribute. \\
    \cline{2-3}
    & $=(v,A)$ & Vector and categorical attribute set.\\
    \hline
    \multicolumn{2}{c|}{$Q=(q,r)$} & Label query vector and its restriction. \\
    \hline
    \multirow{2}{*}{$r$} & $=(l,u)$ & Lower/upper bound for range query. \\
    \cline{2-3}
    & $=f$ & Label for label query. \\
    \hline
    
    \multicolumn{2}{c|}{$n$} & Size of $\mathcal{D}$. \\
    \hline
    \multicolumn{2}{c|}{$\mathcal{D}_r$} & Subset of $\mathcal{D}$ filtered by the restriction $r$. \\
    \hline
    \multicolumn{2}{c|}{$d$} & Dimension per vector in $V$. \\
    \hline
    \multicolumn{2}{c|}{$M$} & Maximum degree for graph ANN index. \\
    \hline
    \multicolumn{2}{c|}{$\phi(v_i,v_j)$} & Similarity between vector $v_i$ and $v_j$. \\
    \hline
  
\end{tabular}
\end{table}



Most Filtering ANN algorithms build upon Inverted File (IVF) or graph-based methods, often enhanced with quantization strategies. To set the stage for filtering methods, we briefly review these foundational ANN approaches, with key notations summarized in Table~\ref{symbol}.

\subsection{Vector Quantization}

Efficient similarity search on high-dimensional data often requires reducing the computational burden. One common approach is to reduce the bit size of each dimension through quantization. For example, Product Quantization (PQ)~\cite{pan2020product, jegou2010product} compresses vectors by clustering one or more dimensions, typically represented as 32-bit floating-point values, into 256 clusters. Each cluster is encoded with an 8-bit code, approximating the original values with substantially lower memory cost.

More advanced quantization techniques, such as Optimized Product Quantization (OPQ)~\cite{ge2013optimized}, Scalar Quantization (SQ)~\cite{dai2021vs}, and RabitQ~\cite{gao2024rabitq}, build on these principles to improve encoding accuracy and computational efficiency, demonstrating strong performance in practical applications.

\subsection{Inverted File}

Another key building block is the Inverted File (IVF) method, which divides vectors into several partitions using k-means clustering~\cite{macqueen1967some}. During a search, only the partitions closest to the query vector are processed.

IVF is commonly combined with quantization techniques (e.g., PQ), forming systems such as IVFPQ~\cite{jegou2010product, douze2024faiss}, to further decrease both computation and memory overhead.


\subsection{Graph-Based ANN Search}

Graph-based ANN search builds an index by connecting vectors based on distance. Key differences among these methods stem from their \textit{pruning strategies} and \textit{entry point} selection. While many approaches exist~\cite{wang2021comprehensive}, only a few are suitable for Filtering ANN. Below, we summarize the most widely used graph indexes in this context.


\begin{figure}[t]
    \centering
    \begin{subfigure}[b]{0.45\linewidth}
        \centering
        \includegraphics[width=0.5\linewidth]{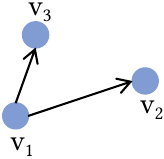}
        \caption{Edge not to be pruned.}
        \label{rngp1}
    \end{subfigure}
    \hfill
    \begin{subfigure}[b]{0.45\linewidth}
        \centering
        \includegraphics[width=0.5\linewidth]{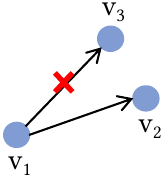}
        \caption{Edge to be pruned.}
        \label{rngp2}
    \end{subfigure}
    \caption{RNG pruning example.}
    \label{RNGprune}
\end{figure}

\textbf{KGraph~\cite{dong2011efficient}} initially connects vectors randomly. For each vector, it iteratively refines the neighbors to the nearest ones by examining the neighbors of its neighbors, following the rule that \textit{neighbors are more likely to be neighbors of each other}~\cite{fu2016efanna}. This design allows KGraph to be constructed efficiently, but it sacrifices connectivity, often resulting in multiple disconnected components~\cite{wang2021comprehensive}.

\textbf{Relative Neighbor Graph (RNG)~\cite{toussaint1980relative}} does not initialize a random graph but instead searches for the nearest neighbors of each vector to be inserted and connects them iteratively. The connection step employs a pruning-based strategy to build a high-quality graph, taking into account the spatial distribution of neighbors. For instance, in Figure~\ref{RNGprune}, suppose point $v_1$ has candidate neighbors $v_2$ and $v_3$, and edge $e(v_1, v_2)$ already exists. RNG prunes edge $e(v_1, v_3)$ if $v_3$ is closer to $v_2$ than to $v_1$, implying that a direct connection is redundant. This strategy maintains a well-connected and scalable structure while limiting the number of neighbors per node. Modern graph-based ANN indexes, such as Vamana Graph~\cite{jayaram2019diskann}, NSG~\cite{fu2017fast}, NSW~\cite{malkov2014approximate}, and HNSW~\cite{malkov2018efficient}, are built upon RNG or its variants.


\textbf{Vamana Graph (VG)~\cite{jayaram2019diskann}}, introduced in DiskANN, starts from a randomly connected graph, performs a one-pass ANN search for each vector, and then prunes each vector’s neighbor list to at most $M$ using the pruning strategy of RNG. This design also retains edges to remote vectors, thereby ensuring connectivity and accelerating search convergence.

\textbf{Hierarchical Navigable Small World (HNSW)~\cite{malkov2018efficient}} constructs a multi-layer graph index where each higher layer is a subsampled version of the one below. In each layer, HNSW leverages the Navigable Small World (NSW) principle by preserving both nearest neighbors for local connectivity and long-range links for global navigation. It also employs an RNG-inspired pruning strategy to eliminate redundant edges, thus ensuring the graph remains sparse yet well-connected. With a single entry point at the top layer, this hybrid design guarantees efficient index construction and rapid query processing.

\begin{table*}[t]
  \caption{Filtering ANN algorithms.}
  \label{overviewFANN}
  \begin{tabular}{c|c|c|c|c}
    \hline
    \multirow{2}{*}{\textbf{Algorithm}} & \textbf{\reviewertwo{Filtering}} & \multirow{2}{*}{\textbf{ANN Index}} & \textbf{Filtering} & \multirow{2}{*}{\textbf{Attribute Index}} \\
    & \textbf{\reviewertwo{Type}} & & \textbf{Strategy} &   \\
    \hline
    \hline
    
    SeRF~\cite{zuo2024serf} & Range & HNSW & Joint-filtering & Segmented edges\\
    \hline
    DSG~\cite{pengdynamic} & Range & HNSW & Joint-filtering & Segmented edges\\
    \hline
    $\beta$-WST~\cite{engels2024approximate} & Range & VG & Pre-filtering & Segmented subgraphs (\reviewerone{Binary search tree})\\
    \hline
    
    UNIFY~\cite{liang2024unify} & Range & HNSW & Pre/Post/Joint-filtering & Segmented subgraphs (\reviewerone{Cross-subgraph edges}) + Skip list\\
    \hline
    iRangeGraph~\cite{xu2024irangegraph} & Range & RNG & Pre-filtering & Segmented subgraphs (\reviewerone{Binary search tree})\\
    \hline
    
    FDiskANN-VG~\cite{gollapudi2023filtered} & Label & VG & Joint-filtering & Labeled edges\\
    \hline
    FDiskANN-SVG~\cite{gollapudi2023filtered} & Label & VG & Joint-filtering & Labeled edges + Stitched graph\\
    \hline
    
    NHQ~\cite{wang2024efficient} & Label & NSW/KGraph & None & Joint distance\\
    \hline

    Milvus~\cite{wang2021milvus} & Arbitrary & HNSW/IVF & Pre-filtering & Partition  \\
    \hline
    Faiss-HNSW~\cite{douze2024faiss, malkov2018efficient} & Arbitrary & HNSW & Post-filtering & Subset identification \\
    \hline
    Faiss-IVFPQ~\cite{douze2024faiss, jegou2010product} & Arbitrary & IVF & Pre-filtering  & Subset identification \\
    \hline
    ACORN~\cite{patel2024acorn} & Arbitrary & HNSW & Joint-filtering & Subset identification + Two-hop scan\\
    \hline

  \hline
\end{tabular}
\end{table*}

\subsection{\reviewertwo{Filtering ANN Search}}

\reviewertwo{Formally, the \textit{Filtering Nearest Neighbor Search (Filtering NN)} is defined as:}
\reviewertwo{\[
NN(q|r) = \arg\min_{o \in D,\, r(o.a)} ||q - o.v||,
\]}
\reviewertwo{where $NN(q|r)$ denotes the exact nearest neighbor whose vector $o.v$ is closest to the query vector $q$ under a given distance metric (e.g., L2, cosine, or inner product), and whose attribute $o.a$ satisfies predicate $r$.}

\reviewertwo{In practice, retrieving exact results is often prohibitive, so approximate solutions (\textit{Filtering ANN}) are used as the algorithmic search target to improve efficiency.}




\section{Overview of Filtering ANN Algorithms}

\label{sec:overview}

\begin{figure}[t]
    \centering
    \includegraphics[width=\linewidth]{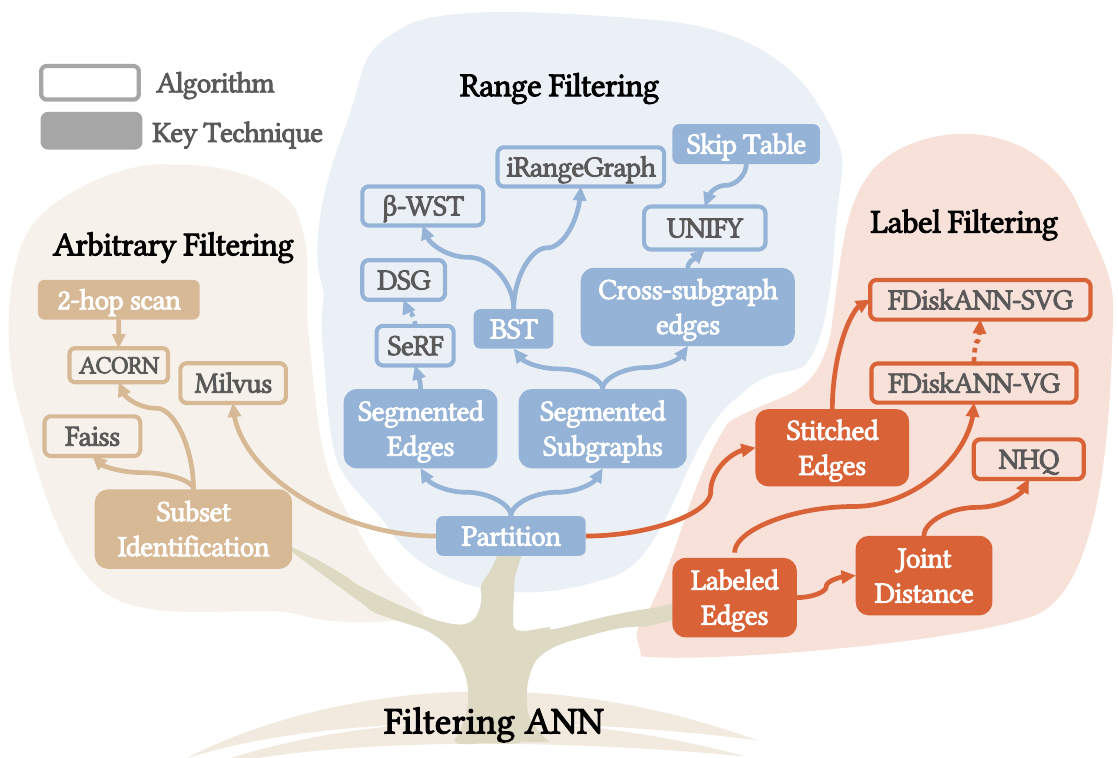}
    \caption{A roadmap of Filtering ANN algorithms.}
    
    \label{roadmap}
\end{figure}

Building on the foundational techniques discussed earlier, recent research has advanced ANN search in filtering scenarios by developing diverse algorithms that combine vector similarity with attribute filtering. Table~\ref{overviewFANN} presents a selection of these methods, which integrate various structural designs and core ideas to achieve distinct performance. 
In the following subsections, we detail approaches for range, label, and arbitrary filtering. Figure~\ref{roadmap} illustrates our taxonomy of the algorithms and their relationship to key techniques.


Index partitioning is employed across all filtering methods, enabling the search range to be pre-defined during construction, especially for range filtering. Label filtering tags edges or encodes labels in distances, while arbitrary filtering identifies subsets before search.

\subsection{Range Filtering ANN Search}

Range Filtering ANN search exploits the natural ordering of continuous numerical attributes, inspiring a diverse set of algorithmic approaches.


\textbf{SeRF~\cite{zuo2024serf}} constructs range-aware edges by first sorting the dataset by attribute values and then incrementally inserting vectors into the graph in ascending order. For each vector \(v_i\), SeRF performs neighbor searches over all subranges \([j, i]\) (with \(j \leq i\)), by creating \textit{segmented edges}. For example, it first identifies the top \(M\) neighbors for the range \([0, i]\), then for \([1, i]\), \([2, i]\), and so on. This segmentation implicitly encodes range constraints via insertion order. During search, SeRF abandons the traditional HNSW structure and instead selects three entry points within the queried range, which enhances efficiency by guiding the search only through valid edges.




\textbf{DSG~\cite{pengdynamic}} builds on SeRF by optimizing construction efficiency. Its key innovation is dynamic vector insertion, which eliminates the need for pre-sorting, thereby enabling real-time index updates.


\textbf{\(\beta\)-WST~\cite{engels2024approximate}} tackles range filtering by constructing a binary search tree (BST) based on attribute ranges, with each BST node hosting a dedicated subgraph that contains only the vectors within that node's attribute range. This hierarchical design yields \(\log(n)\) layers, confining the search to the subgraphs relevant to the query range. In its basic form, a query may intersect multiple subgraphs, potentially reducing efficiency. To address this, an enhanced variant, OptPostFiltering, introduces controlled subgraph overlap combined with post-filtering to streamline the search process.



\textbf{iRangeGraph~\cite{xu2024irangegraph}} also leverages a BST to build \(\log(n)\) layers of subgraphs but differs in query execution. Instead of independently searching subgraphs and merging results, iRangeGraph collects entry points from all subgraphs matching the query range and uses them to traverse the subgraphs as if they are unified, thus avoiding the overhead of post-filtering or result merging.

\textbf{UNIFY~\cite{liang2024unify}} partitions the dataset into attribute-based segments and builds subgraphs on a unified HNSW index. Subgraphs are linked via \textit{Cross-subgraph edges}, with each edge annotated by a \textit{mask list} for segment membership. The search strategy adapts dynamically based on selectivity and the hyperparameters \texttt{Sel\_low} and \texttt{Sel\_high}:

\begin{itemize}[leftmargin=*, nosep]
    \item \ul{Small-range queries (selectivity < \texttt{Sel\_low}):} Uses a \textit{skip list} to pre-filter and scan only the matched vectors.
    \item \ul{Medium-range queries (selectivity \texttt{Sel\_low}$\sim$\texttt{Sel\_high}):} Applies joint-filtering within the pertinent segmented subgraph.
    \item \ul{Large-range queries (selectivity > \texttt{Sel\_high}):} Switches to post-filtering over the full HNSW index.
\end{itemize}

Although coarse segmentation can be challenging for low-selectivity queries, UNIFY’s adaptive strategy selection maintains efficient performance across diverse query types.

\subsection{Label Filtering ANN Search}


Label filtering retrieves similar vectors that satisfy categorical constraints in datasets with a limited number of distinct labels.

\textbf{Filtered-DiskANN~\cite{gollapudi2023filtered}} extends the Vamana Graph (VG)~\cite{jayaram2019diskann} by 
introducing \textit{labeled edges} to navigate query searching across valid nodes.
Filtered-DiskANN introduces the \textit{Stitched graph} to enhance connectivity, which builds subgraphs for individual categorical values and then merges them into a unified graph. Although this approach slows index construction, it significantly improves filtering performance.



\textbf{NHQ~\cite{wang2024efficient}} supports multi-attribute queries where each vector is labeled with exactly one value per attribute (e.g., color, trademark, and origin), and queries must specify a value for every attribute. To combine vector similarity with label matching, NHQ employs a \textit{joint distance} metric defined as:
\[
\phi'(O_i, O_j) = w_1 \cdot \phi(O_i.v, O_j.v) + w_2 \cdot \sum_{t=1}^{|O.A|} \mathbb{I}\{O_i.A_t = O_j.A_t\},
\]
where \(\phi(O_i.v, O_j.v)\) is the vector distance between \(O_i\) and \(O_j\), \(\mathbb{I}\{O_i.A_t = O_j.A_t\}\) is an indicator function that returns 1 if the \(t\)-th attribute of \(O_i\) matches that of \(O_j\) and 0 otherwise, and \(w_1\) and \(w_2\) balance the contributions from the vector and attribute similarities. Although NHQ efficiently guides search using this heuristic, its results may not always perfectly match the filter criteria.

\subsection{Arbitrary Filtering ANN Search}

Arbitrary filtering encompasses scenarios where filtering conditions are flexible, allowing users to define custom subsets of the dataset or custom filtering functions.



\textbf{Faiss~\cite{douze2024faiss}} is a library that supports various ANN methods (e.g., HNSW, LSH, IVF) as well as quantization techniques (e.g., PQ, SQ). It enables arbitrary filtering via 
\textit{Subset identification}, which labels matched items before Filtering ANN search. The \texttt{is\_member(id)} function checks whether a vector belongs to the subset that meets the filtering constraint during search. In Faiss's HNSW implementation, \texttt{is\_member(id)} is invoked during the final layer search to filter results (post-filtering), whereas in IVFPQ it is applied after selecting the \texttt{nprobe} clusters and before computing distances (pre-filtering), thus avoiding unnecessary computations.

\textbf{Milvus~\cite{wang2021milvus}}, built on Faiss, is a vector database that supports arbitrary filtering conditions. 
\reviewertwo{\textsf{Milvus} is popular for being a representative of several vector database systems for its high efficiency, versatility~\cite{pan2024survey}, and widespread adoption in LLM applications~\cite{edge2024local}}.  
It improves filtering performance by first \textit{partitioning} the dataset into multiple subsets (64 by default) based on a specified attribute, which simplifies search tasks confined to specific segments.


\textbf{ACORN~\cite{patel2024acorn}} employs a predicate-agnostic strategy by ignoring filtering constraints during index construction. Its indexing method is similar to HNSW but incorporates two key modifications: it prunes neighbors using a \textit{two-hop} rule (see Section~\ref{pruning}), and, during the search, it expands to two-hop neighbors when immediate neighbors are insufficient. While effective for arbitrary filtering, this general-purpose connectivity may lead to slower convergence for categorical or numerical queries compared to specialized methods.





\section{Detailed Analysis of Key Components}

\label{sec:component}

In this section, we examine the core techniques and design components of Filtering ANN algorithms, emphasizing their distinctive features and implementation nuances.

\subsection{Attribute Index}

We focus on analyzing range Filtering ANN indexes, as they are significantly more complex and worthy of detailed examination. 
Range filtering relies on partitioning the dataset according to the natural ordering of numerical attributes. As used in Milvus, a simple strategy is to divide the dataset into segments and build a separate index for each subset. However, many range filtering algorithms employ more sophisticated segmentation strategies. Based on the techniques summarized in Table~\ref{overviewFANN} and Figure~\ref{roadmap}, we classify them into two main approaches: \textbf{segmented edges} and \textbf{segmented subgraphs}.

\begin{figure}[t]
    \centering
    \hfill
    \begin{subfigure}[b]{0.5\linewidth}
        \centering
        \includegraphics[width=\linewidth]{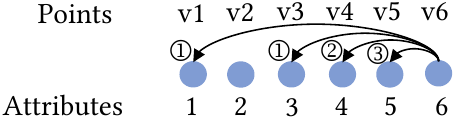}
        \caption{Search range selection steps}
        \label{serfconnect}
    \end{subfigure}
    \hfill
    \begin{subfigure}[b]{0.45\linewidth}
        \centering
        \includegraphics[width=\linewidth]{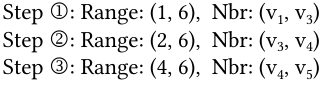}
        \caption{Edge selection steps}
        \label{serfstep}
    \end{subfigure}
    \caption{Segmented edge selection example. M = 2; the order of distances to $v_6$ is: $v_1$, $v_3$, $v_4$, $v_5$, $v_2$; `Nbr' denotes the selected neighbors for each range.}
    \label{serfedgesel}
\end{figure}

\ul{\textbf{Segmented edges.}} SeRF and DSG build edges annotated with range indicators that support queries over different intervals, ensuring that every query can effectively locate its target vectors. The core idea is to create edges covering all possible query ranges. 
Figure~\ref{serfconnect} illustrates how the new node \(v_6\) is inserted and connected to existing nodes, while Figure~\ref{serfstep} presents its edge sets along with their corresponding filtering ranges.
The algorithm performs a progressive search, from the largest to the smallest query range, as follows:

\begin{itemize}
    \item[\(\textcircled{1}\).] For range \((1,6)\), select the nearest neighbors \(v_1\) and \(v_3\).
    \item[\(\textcircled{2}\).] For range \((2,6)\), choose \(\{v_3, v_4\}\), reusing distances computed in \(\textcircled{1}\).
    \item[\(\textcircled{3}\).] Skip range \((3,6)\) as it shares the same neighbors as $(2,6)$; for range \((4,6)\), select \(v_4\) and \(v_5\).
\end{itemize}


Technically, DSG constructs one extra range for each edge, ensuring that they can guide range queries that do not match the current node and help find closer paths. Additionally, incremental insertion is supported, as the insertion order is no longer a strict requirement.


\begin{figure}[t]
    \centering
    \hfill
    \begin{subfigure}[b]{0.4\linewidth}
        \centering
        \includegraphics[width=\linewidth]{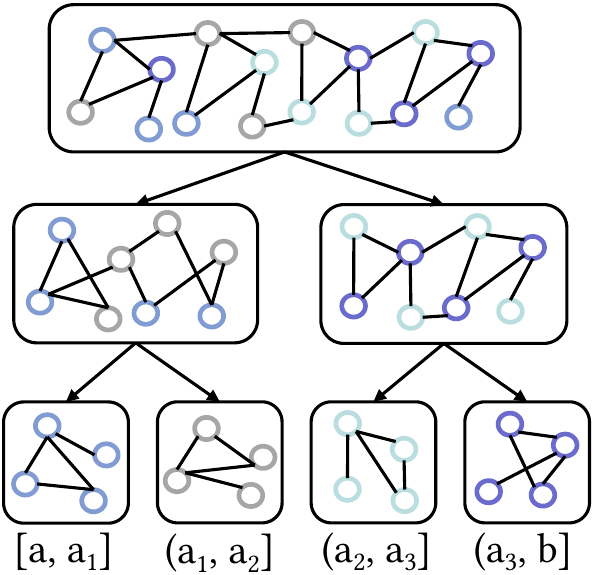}
        \caption{Binary search tree}
        \label{bst}
    \end{subfigure}
    \hfill
    \begin{subfigure}[b]{0.45\linewidth}
        \centering
        \includegraphics[width=\linewidth]{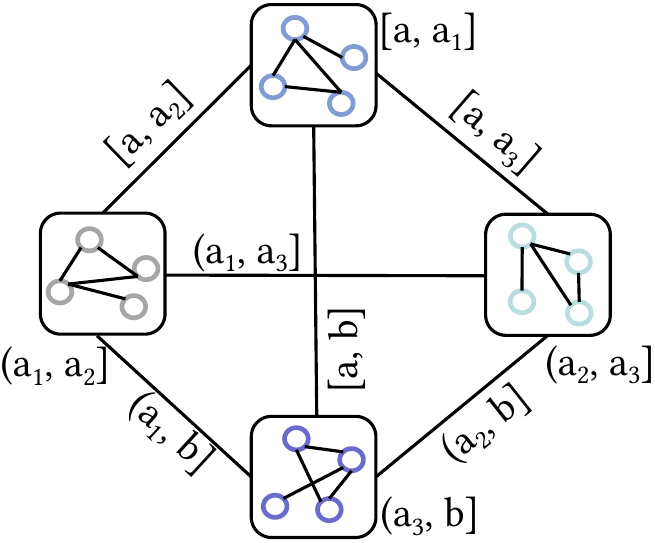}
        \caption{Cross-subgraph edges}
        \label{cross}
    \end{subfigure}
    \caption{Different implementations of segmented subgraph approaches.}
    \label{subgraph}
\end{figure}

\ul{\textbf{Segmented subgraphs.}} 
Another strategy for range Filtering ANN indexing is to build dedicated subgraphs for different query intervals. For instance, \(\beta\)-WST and iRangeGraph organize data into a BST, with each tree node hosting a subgraph that covers a specific range, as illustrated in Figure~\ref{bst}. Queries then combine the relevant subgraphs from the corresponding BST layers. UNIFY adopts a similar approach but creates a fixed set of disjoint subgraphs—one per range—and introduces Cross-subgraph edges  to maintain global connectivity, as shown in Figure~\ref{cross}.

A BST can produce up to \(\log(n)\) layers, resulting in \(2n-1\) subgraphs and supporting a large spectrum of range combinations. By contrast, UNIFY’s subgraph count remains fixed, generating fewer possible combinations. Consequently, UNIFY relies on post-filtering for specialized or narrow ranges, reducing its efficiency under low-selectivity conditions.

\subsection{Pruning Techniques}
\label{pruning}

Many Filtering ANN algorithms (e.g., Milvus, SeRF, DSG, $\beta$-WST, iRangeGraph, UNIFY) employ RNG pruning to boost search efficiency. For each vector, RNG pruning removes edges to vectors that are too close to its current neighbors, retaining only distinctive, diverse connections. However, since RNG pruning considers only distance and ignores attribute diversity, a crucial factor in filtering scenarios, several algorithms have introduced specialized pruning techniques.


    
\begin{figure}[t]
    \centering
    \begin{subfigure}[b]{0.23\linewidth}
        \centering
        \includegraphics[width=0.7\linewidth]{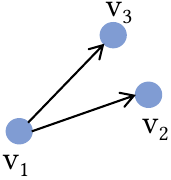}
        \caption{}
        \label{acornnp}
    \end{subfigure}
    \hfill
    \begin{subfigure}[b]{0.22\linewidth}
        \centering
        \includegraphics[width=0.7\linewidth]{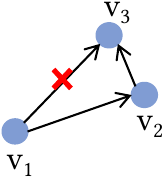}
        \caption{}
        \label{acornp}
    \end{subfigure}
    \hfill
    \begin{subfigure}[b]{0.25\linewidth}
        \centering
        \includegraphics[width=0.85\linewidth]{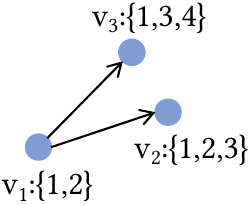}
        \caption{}
        \label{fdisknp}
    \end{subfigure}
    \hfill
    \begin{subfigure}[b]{0.23\linewidth}
        \centering
        \includegraphics[width=\linewidth]{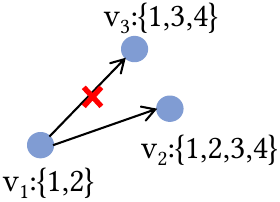}
        \caption{}
        \label{fdiskp2}
    \end{subfigure}

    \caption{Examples of two-hop and label pruning: (a) Edge not to be pruned in two-hop pruning, (b) Edge to be pruned in two-hop pruning, (c) Edge not to be pruned in label-covered pruning, and (d) Edge to be pruned in label-covered pruning.} 
    \label{filteredprune}
\end{figure}

\ul{\textbf{Two-hop pruning.}} ACORN restricts pruning to \emph{two-hop} neighbors (Figure~\ref{acornp}) without caring about relative distances like RNG but aligning with its two-hop {\em search} strategy. By excluding immediate neighbors from pruning, ACORN retains more potentially useful connections, improving both search efficiency and accuracy.

\ul{\textbf{Label-covered pruning.}}  
Filtered-DiskANN enhances RNG pruning by integrating label information. An edge \(e(v_1, v_3)\) is pruned only if two conditions hold: (1) the RNG condition is met—that is, there exists a vector \(v_2\) such that \(\phi(v_1, v_2) < \phi(v_1, v_3)\) and \(\phi(v_3, v_2) < \phi(v_1, v_3)\); and (2) \(v_2\)'s attribute set covers those of both \(v_1\) and \(v_3\). For instance, Figure~\ref{fdiskp2} shows that \(e(v_1, v_3)\) is pruned when both conditions are satisfied, whereas in Figure~\ref{fdisknp} only the RNG condition holds, so the edge is retained. This approach ensures that edges are pruned only when both spatial proximity and label consistency warrant it, thereby preserving crucial connectivity for categorical filtering tasks.


\subsection{Entry Point Strategies}


Graph-based Filtering ANN search must eliminate mismatched neighbors at various stages, and a critical challenge is selecting reliable entry points for graph traversal. Although Table~\ref{overviewFANN} provides a general classification, important implementation nuances warrant further discussion.

\ul{\textbf{Unrestricted entry points.}}  
Traditional approaches—such as Faiss-HNSW with \texttt{is\_member(id)} restrictions—preserve the original entry points for queries. Similarly, post-filtering methods in ACORN, NHQ, and UNIFY initiate searches from default entry points of the graph index, gradually converging on matching results.


\ul{\textbf{Specialized entry points.}}
SeRF and DSG bypass the hierarchical navigation of HNSW by directly selecting multiple entry points from the bottom layer that satisfy the query range constraints. Instead of using a hierarchical scheme, they select entry points at evenly spaced intervals within the valid range, ensuring a well-distributed set of starting points without extra overhead.

Other methods handle entry point selection differently. For example, iRangeGraph dynamically combines entry points from multiple matched subgraphs during query processing, while UNIFY’s joint-filtering uses a single top-level entry point from selected subgraphs and connects them via cross-subset edges. In its pre-filtering mode, UNIFY begins from a fixed entry point and delays distance computations until the skip table reaches the query’s left bound. 






\section{Experiments}

In this section, we present our experimental setup and results, offering a comprehensive evaluation of various Filtering ANN algorithms along with an in-depth analysis of their key components.


\subsection{Setup} 
\label{sec:exp:setup}

\textbf{Platform.} We conducted our experiments on Ubuntu 24.04 LTS, equipped with Intel\textsuperscript{\textregistered} Xeon\textsuperscript{\textregistered} Platinum 8358 CPUs @ 2.60GHz, x86-64 architecture, and 2TB of memory, \reviewerone{with 128 physical cores. We use 128 threads for index construction. As query tasks are read-only, all methods exhibit similar performance trends with increased parallelism, so we use one thread during the query phase.}

\textbf{Metrics.} All datasets use L2 (Euclidean) distance for similarity measurement. We evaluate algorithms using \texttt{Queries Per Second (QPS)}, where a higher QPS indicates faster processing, and \reviewerone{\texttt{Comparisons}} per query for graph-based algorithms, where a fewer comparisons signifies more efficient navigation. \reviewerthree{Since all methods apply similar SIMD-based techniques for distance computation, \texttt{QPS} and \texttt{comparisons} exhibit consistent trends. } 


\textbf{Dataset.} Our experiments are conducted on four datasets: SIFT \cite{jegou2011searching, aumuller2020ann}, SpaceV \cite{benchmark2021billion}, Redcaps \cite{desai2021redcaps}, and Youtube-RGB (see Table~\ref{dataset}). These datasets include up to 10 million vectors with varying dimensions and label types (synthetic or real). Unless specified otherwise, real labels are used for Redcaps and Youtube-RGB.

\begin{table}[t]
  \caption{Dataset statistics.}
  \label{dataset}
  \begin{tabular}{c|c|c|c|c}
    \hline
    \textbf{Dataset} & \textbf{Dim} & \textbf{Labels} & \textbf{Size} & \textbf{Query Size}\\
    \hline
    \hline
    SIFT \tablefootnote{http://corpus-texmex.irisa.fr/} & 128 & Synthetic & 10M & 10K \\
    \hline
    SpaceV \tablefootnote{https://github.com/microsoft/SPTAG/tree/main/datasets/SPACEV1B} & 100 & Synthetic & 10M & 10K \\
    \hline
    Redcaps \tablefootnote{https://redcaps.xyz/} & 512 & Real/Synthetic & 1M & 10K \\
    \hline
    Youtube-RGB\tablefootnote{https://research.google.com/youtube8m/download.html} & 1024 & Real/Synthetic & 1M & 10K \\
    \hline
\end{tabular}
\end{table}
\vspace{-2pt}

To ensure a fair comparison across all algorithms, synthetic labels are generated using the following strategy: each vector is assigned an integer value between 0 and 100,000 for numerical attributes, \reviewerone{and one of 500 integer values for categorical attributes, maintaining consistent label cardinality for both Filtered-DiskANN and NHQ. For each query task, we retrieve 10,000 items sequentially to evaluate the performance of all algorithms. For range queries, we control selectivity by adjusting the upper and lower bounds of the queried attribute. For label queries, we assign categorical attributes using fixed probabilities to control selectivity (e.g., assigning Label 1 to 50\% of vectors so that querying with Label 1 yields 50\% selectivity).}

\begin{figure}[t]
  \centering
  \includegraphics[width=\linewidth]{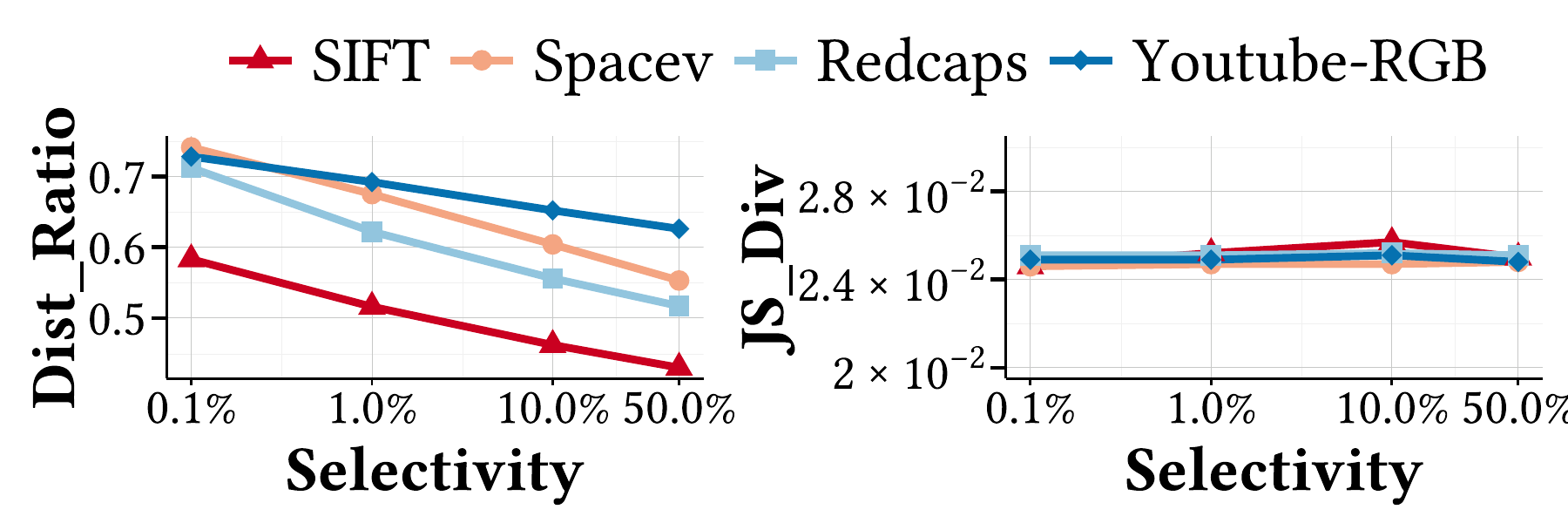}
  \caption{\reviewerthree{Query hardness of all datasets.}}
  \label{fig:dataset_dist}
\end{figure}

\reviewerthree{Figure~\ref{fig:dataset_dist} analyzes dataset hardness under varying selectivity levels. We define Dis\_Ratio as the ratio between the average distance of the top-10 ground truth and the average pairwise distance in the dataset; a higher value implies a broader search range. We also estimate the Jensen-Shannon divergence (JS\_Div) between the full dataset and the queried subset. In general, lower selectivity leads to harder queries, and both synthetic and real datasets exhibit limited divergence in query distribution.}

\textbf{Algorithms.}  
In our experiments, we evaluate the latest Filtering ANN algorithms, grouped into three categories:
\begin{sloppypar}
\begin{itemize}
    \item \textbf{Range Filtering Methods:}
    \begin{itemize}
         \item \textsf{SeRF~\cite{zuo2024serf}} and \textsf{DSG~\cite{pengdynamic}}
         \item \textsf{$\beta$-WST~\cite{engels2024approximate}}, evaluated in two modes: the Vamana graph filtering method (\textsf{WST-Vamana}) and a super-optimized post-filtering variant (\textsf{WST-opt})
         \item \textsf{iRangeGraph~\cite{xu2024irangegraph}} and \textsf{UNIFY~\cite{liang2024unify}}, evaluated in two configurations: a hybrid pre-/post-/joint-filtering strategy (\textsf{UNIFY-CBO}) and a joint-filtering-only setup (\textsf{UNIFY-joint})
    \end{itemize}

    \item \textbf{Label Filtering Methods:}
    \begin{itemize}
         \item \textsf{Filtered-DiskANN~\cite{gollapudi2023filtered}}, evaluated in both its base (\textsf{FDiskANN-VG}) and stitched (\textsf{FDiskANN-SVG}) forms
         \item \textsf{NHQ~\cite{wang2024efficient}}, implemented with NSW (\textsf{NHQ-NSW}) and KGraph (\textsf{NHQ-KGraph})
    \end{itemize}

    \item \textbf{Arbitrary Filtering Methods:}
    \begin{itemize}
         \item \textsf{Faiss~\cite{douze2024faiss}} (\textsf{Faiss-HNSW}, \textsf{Faiss-IVFPQ})
         \item \textsf{Milvus~\cite{wang2021milvus}} (\textsf{Milvus-HNSW}, \textsf{Milvus-IVFPQ})
         \item \textsf{ACORN~\cite{patel2024acorn}}
    \end{itemize}
\end{itemize}
\end{sloppypar}


To ensure fairness, we standardize key settings: parallelism is enabled for \textsf{SeRF} and \textsf{DSG}, \textsf{ACORN}'s filtering storage is optimized, and \textsf{Faiss}'s \texttt{is\_member()} is improved. \reviewerone{All methods use in-memory indexes, including \textsf{DiskANN}, which has a built-in in-memory search component.} All graph-based methods use \(M = 40\) and \texttt{ef\_construction} = 1000. \reviewerthree{ These settings yield near-optimal performance for up to 10M datasets~\cite{pengdynamic, malkov2018efficient}.} \reviewerone{Each experiment is averaged over three runs}. Full configurations are in the appendix~\cite{appendix}.

\subsection{Range Filtering: Performance and Analysis}

\label{sec:exp:range}

\begin{figure*}[htbp]
	\centering
	\begin{subfigure}{\linewidth}
		\centering
		\includegraphics[width=\linewidth]{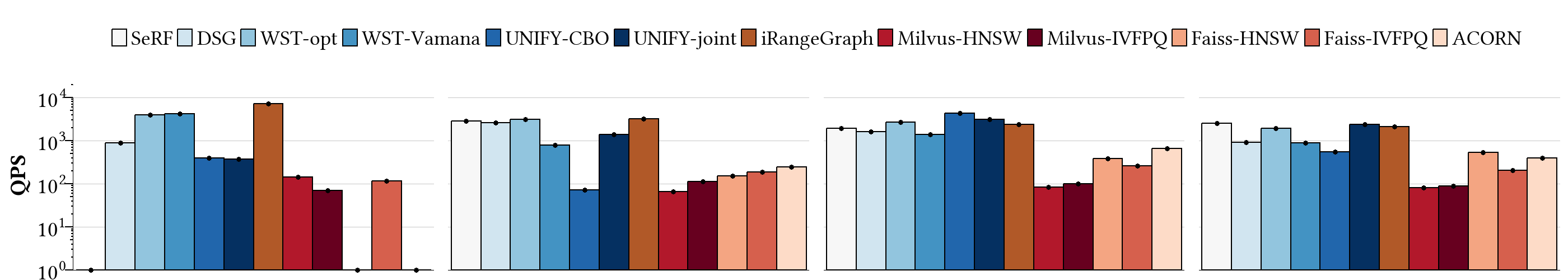}
		\caption{SIFT}
		\label{sift}
	\end{subfigure}
    \\
	\centering
	\begin{subfigure}{\linewidth}
		\centering
		\includegraphics[width=\linewidth]{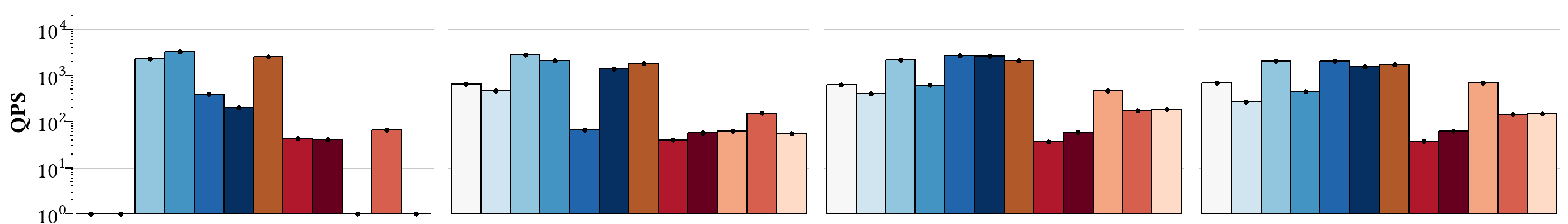}
		\caption{Spacev}
		\label{spacev}
	\end{subfigure}
    \\
	\centering
	\begin{subfigure}{\linewidth}
		\centering
		\includegraphics[width=\linewidth]{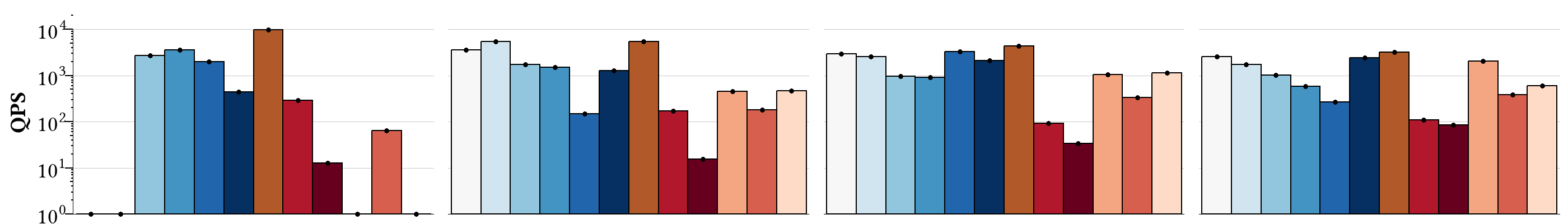}
		\caption{Redcaps}
		\label{redcaps}
	\end{subfigure}
    \\
	\centering
	\begin{subfigure}{\linewidth}
		\centering
		\includegraphics[width=\linewidth]{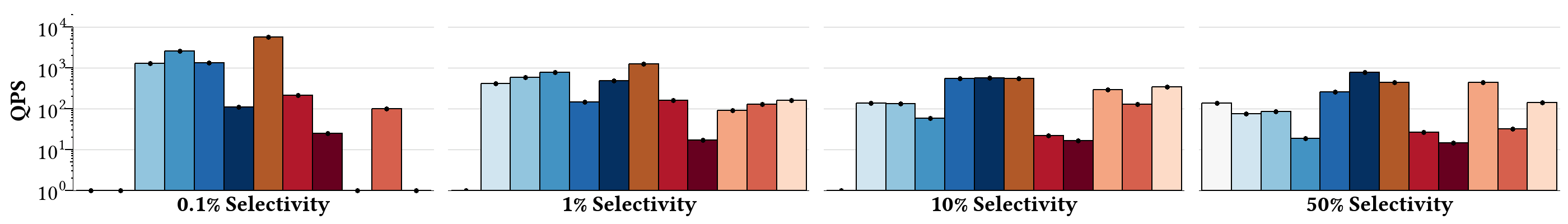}
		\caption{Youtube-RGB}
		\label{youtube}
	\end{subfigure}
	\caption{QPS for range Filtering ANN algorithms at 90\% recall\text{@}10.}
	\label{qps}
\end{figure*}

\begin{figure*}[t]
	\centering
	\begin{subfigure}{\linewidth}
		\centering
		\includegraphics[width=\linewidth]{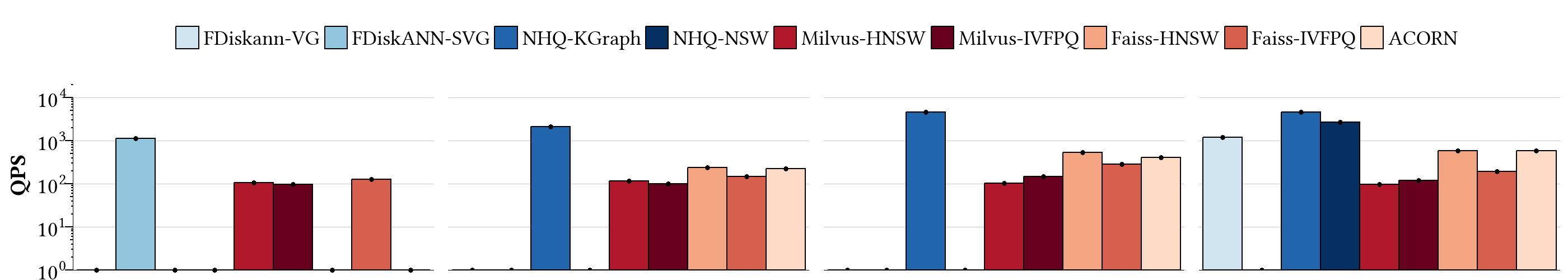}
		\caption{SIFT}
		\label{sift2}
	\end{subfigure}
    \\
	\centering
	\begin{subfigure}{\linewidth}
		\centering
		\includegraphics[width=\linewidth]{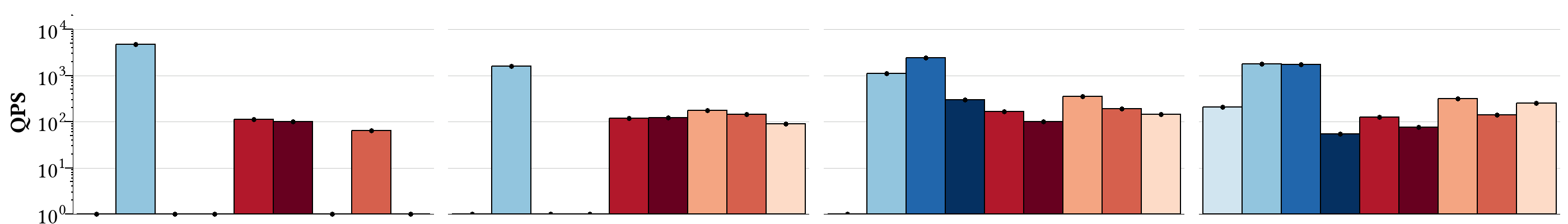}
		\caption{Spacev}
		\label{spacev2}
	\end{subfigure}
    \\
	\centering
	\begin{subfigure}{\linewidth}
		\centering
		\includegraphics[width=\linewidth]{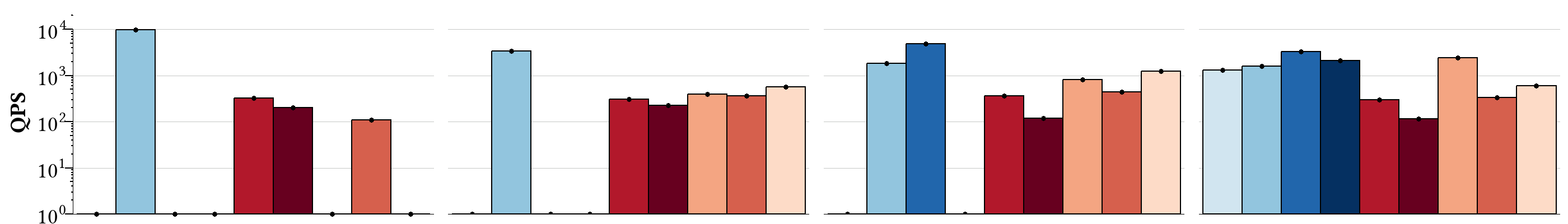}
		\caption{Redcaps-Synthetic}
		\label{redcaps2}
	\end{subfigure}
    \\
	\centering
	\begin{subfigure}{\linewidth}
		\centering
		\includegraphics[width=\linewidth]{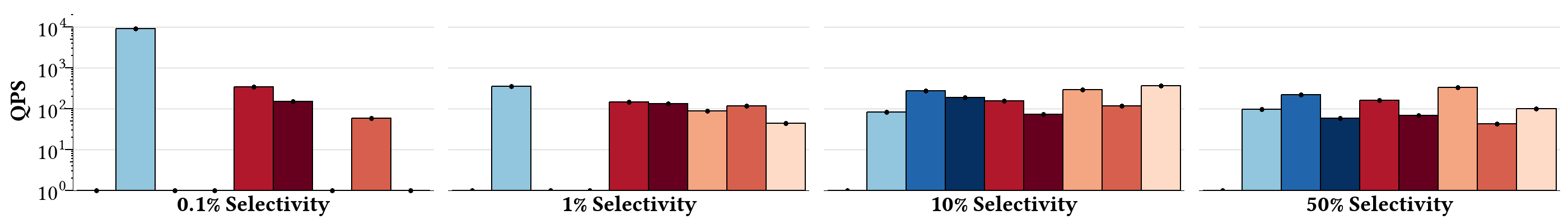}
		\caption{Youtube-RGB-Synthetic}
		\label{youtube2}
	\end{subfigure}
	\caption{QPS for label Filtering ANN algorithms at 90\% recall\text{@}10.}
	\label{qpslabel}
\end{figure*}

We evaluate algorithm performance over selectivity levels of 0.1\%, 1\%, 10\%, and 50\% to assess efficiency under various query scenarios. 
\reviewerone{Numerical attributes are generated based on the rules in Section~\ref{sec:exp:setup}. We fine-tune search hyper-parameters (e.g., \texttt{ef\_search}, \texttt{nprobe}, etc.) to achieve optimal performance at 90\% recall. }
Results are shown in Fig. \ref{qps}. 
Missing entries indicate failure to meet the target recall under equivalent indexing configurations. Our key observations are as follows:

\ul{\textbf{1. Partitioning ensures query availability at low selectivity.}} \textsf{Milvus} achieves reliable query service at 0.1\% selectivity, a challenging regime where \textsf{Faiss-HNSW} fails completely. This capability stems from its partitioning mechanism: by distributing the dataset across smaller subsets based on attributes, the selectivity within each partition becomes significantly higher than the global selectivity.

\ul{\textbf{2. IVF-based methods show reliable recall.}} 
Although \textsf{Faiss-IVFPQ} generally achieves lower QPS than \textsf{Faiss-HNSW}, it consistently handles 0.1\% selectivity across all datasets. In contrast, \textsf{Faiss-HNSW} with post-filtering and \textsf{ACORN} often fail under these conditions. This discrepancy is due to the monotonic search behavior of \textsf{HNSW}, which restricts exploration as the search converges on the target, whereas the non-monotonic nature of \textsf{IVF} enables it to explore a broader search range and reliably locate matching targets.


\ul{\textbf{3. The benefits of attribute-aware indexing are less pronounced at high selectivity levels.}}  
Our experiments reveal that when selectivity exceeds 50\%, even traditional methods like \textsf{Faiss-HNSW} achieve competitive performance. This is expected, as the benefit of attribute-aware indexing is most pronounced under stringent filtering conditions; at high selectivity, the search range is broad enough that the extra filtering constraints provide little advantage.

\ul{\textbf{4. Range filtering methods achieve similar performance in most cases.}} 
\textsf{UNIFY-hybrid}, \textsf{WST-opt}, \textsf{iRangeGraph}, \textsf{SeRF} and \textsf{DSG} consistently yield the highest QPS in most test scenarios (1\% and 10\% selectivity), thanks to subgraph constructions that precompute edges across all query ranges, ensuring robust connectivity during search. This demonstrates that segmented edges and subgraphs offer similar effectiveness for range filtering.

\ul{\textbf{5. BST excels at low selectivity compared to cross-subgraph edges.}}
\textsf{iRangeGraph} and \textsf{WST-Vamana} employ fine-grained subgraphs constructed using BST, achieving outstanding performance at 0.1\% selectivity. In contrast, \textsf{UNIFY-hybrid}’s default partitioning into 8 subsets performs well at higher selectivity levels but suffers from reduced QPS at 0.1\%, even falling behind \textsf{UNIFY-CBO}’s linear scan based on the skip table.



\ul{\textbf{6. Segmented edges fail at low selectivity.}} 
Although \textsf{SeRF} and \textsf{DSG} deliver high QPS at high selectivity, both methods fail at 0.1\% selectivity. Because they rely on a graph-based structure, this issue mirrors the monotonicity challenge found in \textsf{Faiss-HNSW}. Under extremely narrow filtering ranges, the initial search during index construction fails due to the absence of valid neighbors, leading subsequent queries to collapse. Notably, \textsf{DSG} underperforms compared to \textsf{SeRF}, primarily because it dedicates significant time to edge filtering despite its more refined strategy. Section~\ref{sec:exp:edgesel} discusses this issue in detail.

\subsection{Label Filtering: Performance and Analysis}

\label{sec:exp:label}

\reviewerone{Following the attribute generation method in Section~\ref{sec:exp:setup} and experimental settings in Section~\ref{sec:exp:range}, we conduct experiments for label filtering.} Figure~\ref{qpslabel} presents the QPS achieved by each label filtering algorithm while maintaining 90\% recall across various selectivity levels. Our analysis reveals several key insights:

\ul{\textbf{1. Robustness of the stitched method.}}  
\textsf{FDiskANN-SVG} generally performs well across datasets and selectivity levels due to its stitched subgraph design. However, it fails to achieve 90\% recall on SIFT, likely owing to limitations in the quality of the underlying Vamana Graph. In contrast, on SpaceV, Redcaps, and Youtube-RGB, \textsf{FDiskANN-SVG} meets or exceeds the recall target, indicating its effectiveness is dataset-dependent. 

\ul{\textbf{2. Limitations of joint distance at low selectivity.}}  
Both \textsf{NHQ-KGraph} and \textsf{NHQ-NSW} suffer substantial performance drops when selectivity falls below 1\%. This decline stems from difficulties in maintaining graph connectivity under the joint distance metric; when nodes sharing the same label are sparse, the metric fails to establish valid neighbor connections, limiting its effectiveness for high-precision filtering.

\ul{\textbf{3. Simpler pruning outperforms in label filtering search.}} 
Our comparison between \textsf{NHQ-KGraph} and \textsf{NHQ-NSW} reveals that simpler pruning strategies can yield superior performance in filtered search. By retaining all nearest neighbors, \textsf{NHQ-KGraph} preserves more connections, leading to improved recall. These findings suggest that, in Filtering ANN contexts, straightforward connectivity rules can outperform more complex geometric pruning techniques.






\subsection{\reviewertwo{Arbitrary Filtering Analysis}}
\begin{figure}[t]
	\centering
	\begin{subfigure}{\linewidth}
		\centering
		\includegraphics[width=\linewidth]{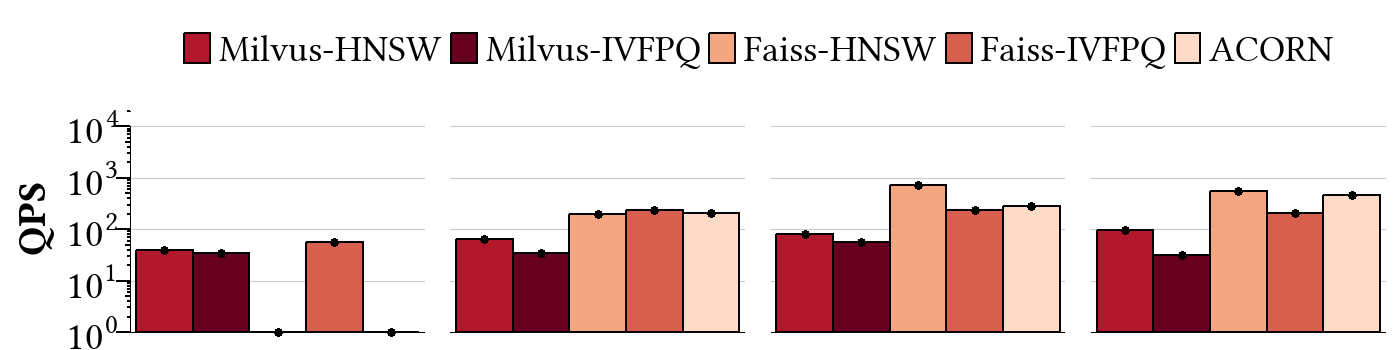}
		\caption{SIFT}
		\label{sift2}
	\end{subfigure}
    \\
	\centering
	\begin{subfigure}{\linewidth}
		\centering
		\includegraphics[width=\linewidth]{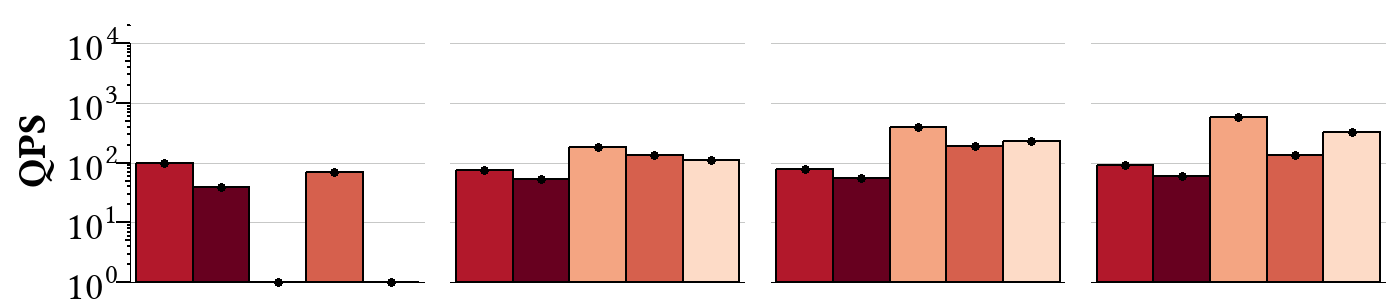}
		\caption{Spacev}
		\label{spacev2}
	\end{subfigure}
    \\
	\centering
	\begin{subfigure}{\linewidth}
		\centering
		\includegraphics[width=\linewidth]{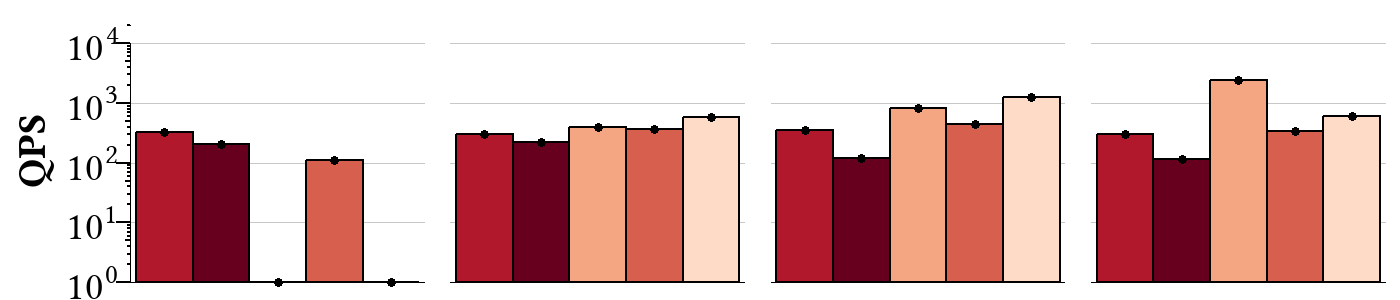}
		\caption{Redcaps-Synthetic}
		\label{redcaps2}
	\end{subfigure}
    \\
	\centering
	\begin{subfigure}{\linewidth}
		\centering
		\includegraphics[width=\linewidth]{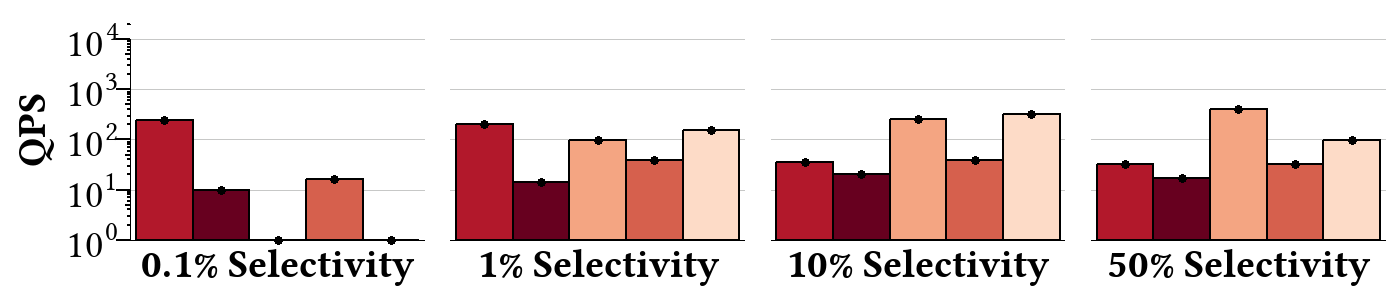}
		\caption{Youtube-RGB-Synthetic}
		\label{youtube2}
	\end{subfigure}
	\caption{\reviewertwo{QPS for arbitrary Filtering ANN algorithms at 90\% recall\text{@}10.}}
	\label{qpsarbi}
\end{figure}


\reviewertwo{To evaluate arbitrary filtering performance, we perform two-predicate searches using both categorical and numerical attributes for hybrid filtering. To achieve 50\% overall selectivity, we assign the categorical label ``1'' to 60\% of vectors and apply a range query with 83\% selectivity. Under uniform distributions, the combined selectivity is approximately $60\% \times 83\% \approx 50\%$. Other selectivity levels are computed similarly. Since \textsf{Milvus} supports partitioning by only one attribute, we use the numerical attribute for partitioning. The results are presented in Figure~\ref{qpsarbi}.}

\reviewertwo{\ul{\textbf{1. \textsf{Faiss-HNSW} and \textsf{ACORN} show competitive performance at high selectivity.}}  
At selectivities greater than 10\%, \textsf{ACORN} and \textsf{Faiss-HNSW} achieve higher QPS due to their single-index design without partitioning. Notably, \textsf{Faiss-HNSW} performs best when selectivity exceeds 50\%. These results confirm that \textsf{Faiss-HNSW} remains a strong candidate for arbitrary filtering.}

\reviewertwo{\ul{\textbf{2. Scanning within a partitioned subset offers a simple yet effective strategy at low selectivity.}}  
\textsf{Milvus-HNSW} demonstrates strong performance at low selectivity levels. Interestingly, varying \textit{ef\_search} has no impact on its QPS or recall, suggesting that \textsf{Milvus-HNSW} performs direct scans within partitioned subsets rather than traversing the graph. In contrast, \textsf{Faiss-HNSW} and \textsf{ACORN} struggle at 0.1\% selectivity, likely due to difficulties in identifying and reaching sparse candidates.}

\reviewertwo{\ul{\textbf{3. Multiple predicate filtering performs similarly to single predicate under uniform distribution.}}  
For uniformly distributed attributes, \textsf{Faiss} and \textsf{ACORN} handle multiple filtering predicates similarly, resulting in comparable performance between single and multiple predicates at the same overall selectivity. However, \textsf{Milvus-HNSW} exhibits different behavior (see point~2) due to changes in its internal search strategy when multiple predicates are applied.}

\subsection{Index Analysis}

\label{sec:exp:index}

\definecolor{bad1}{HTML}{ca0020}
\definecolor{good1}{HTML}{0571b0}
\begin{table*}[t]

  \caption{Index size and index time.}
  \label{tableindex}
  \begin{tabular}{c|c|c|c|c|c|c|c|c|c|c|c|c|c}
  
\hline
\textbf{Filtering}&
\multirow{2}{*}{\textbf{Algorithm}} & \multicolumn{3}{c|}{ \textbf{SIFT (4.80GB)} } & \multicolumn{3}{c|}{ \textbf{spacev (3.76GB)} } & \multicolumn{3}{c|}{ \textbf{Redcaps (1.91GB)} } & \multicolumn{3}{c}{ \textbf{Youtube-RGB (3.81GB)} }\\
\cline{3-14}
 \textbf{Method} & & \reviewerthree{Size} &  Mem & Time & \reviewerthree{Size} & Mem & Time  & \reviewerthree{Size} & Mem & Time  & \reviewerthree{Size} & Mem & Time \\
\hline
\hline
\multirow{5}{*}{Arbi-} & \textsf{Faiss-HNSW}  & 7.90 & 7.99 & 1733 & 6.86 & 6.93 & 2564 & 2.22 & 2.26 & 247 & 4.13 & 4.18 & 380\\
& \textsf{Faiss-IVFPQ}  & \cellcolor{good1!25}0.68 & \cellcolor{good1!25}2.90  &  1354 & \cellcolor{good1!25}0.54 & \cellcolor{good1!25}2.28  &  1370 & \cellcolor{good1!25}0.25 & \cellcolor{good1!25}1.27  &  \cellcolor{bad1!25}1466 & \cellcolor{good1!25}0.50 &  \cellcolor{good1!25}2.51 &  \cellcolor{bad1!25}2165\\
& \textsf{Milvus-HNSW}  & - & 49.15  &  661 & - & 49.15  &  939 & - & \cellcolor{bad1!25}49.15  &  \cellcolor{good1!25}235 & - & \cellcolor{bad1!25}49.15  & \cellcolor{good1!25}311\\
& \textsf{Milvus-IVFPQ}  & - & 49.15  &  \cellcolor{good1!25}507 & - & 49.15  &  \cellcolor{good1!25}621 & - & \cellcolor{bad1!25}49.15  &  323 & - & \cellcolor{bad1!25}49.15  &  548\\
& ACORN & \cellcolor{bad1!25}10.45 & \cellcolor{bad1!25}109.33 &  \cellcolor{bad1!25}3859 & \cellcolor{bad1!25}9.41 & \cellcolor{bad1!25}108.29 &  \cellcolor{bad1!25}3321 &  \cellcolor{bad1!25}2.48 & 12.39 & 810 & \cellcolor{bad1!25}4.38 & 14.31  &  757\\
\hline
\multirow{4}{*}{Label-} & \textsf{FDiskANN-VG} & 6.34 & 7.24  &  505 & \cellcolor{bad1!25}5.30 & 6.34 & 465 & \cellcolor{bad1!25}2.07 & 2.19 & 35 & \cellcolor{bad1!25}3.97 & 4.12  &  \cellcolor{good1!25}31\\
& \textsf{FDiskANN-SVG} & \cellcolor{bad1!25}6.35 & 7.11  &  370 & \cellcolor{bad1!25}5.30 & 6.27  &  325 & \cellcolor{bad1!25}2.07 & 2.16  &  \cellcolor{bad1!25}210 & \cellcolor{bad1!25}3.97 & \cellcolor{good1!25}4.00  &  100\\
& \textsf{NHQ-KGraph} & \cellcolor{good1!25}5.71 & \cellcolor{bad1!25}12.90  &  \cellcolor{good1!25}266 & \cellcolor{good1!25}4.67 & \cellcolor{bad1!25}11.06  &  \cellcolor{good1!25}252 & \cellcolor{good1!25}1.97 & \cellcolor{bad1!25}4.14  &  \cellcolor{good1!25}17 & \cellcolor{good1!25}3.83 & \cellcolor{bad1!25}7.94  &  120\\
& \textsf{NHQ-NSW}  & 6.30  & \cellcolor{good1!25}6.36  &  \cellcolor{bad1!25}1658 & 5.26 & \cellcolor{good1!25}5.31  &  \cellcolor{bad1!25}1699 & 2.06 & \cellcolor{good1!25}2.11  &  171 & \cellcolor{bad1!25}3.97 & 4.04  &  \cellcolor{bad1!25}367\\
\hline
\multirow{6}{*}{Range-}& \textsf{SeRF} & \cellcolor{good1!25}6.96  & \cellcolor{good1!25}10.44  &  \cellcolor{good1!25}364 & \cellcolor{good1!25}5.88 & \cellcolor{good1!25}8.35  &  \cellcolor{good1!25}374 & \cellcolor{good1!25}2.09 & \cellcolor{good1!25}3.93  &  \cellcolor{good1!25}61 & \cellcolor{good1!25}3.89  & 7.76 & \cellcolor{good1!25}45\\
& \textsf{DSG} & 39.35 & 40.35  &  2905 & \cellcolor{bad1!25}36.88 & 37.87  &  3411 & \cellcolor{bad1!25}5.09 & 5.22  &  358 & \cellcolor{bad1!25}5.10 & 7.76  &  1045\\
& \textsf{WST-opt}  & \cellcolor{bad1!25}40.22 & \cellcolor{bad1!25}97.69  &  8048 & 34.68 & \cellcolor{bad1!25}92.66  &  8775 & 3.69 & \cellcolor{bad1!25}11.70  &  2391 & 4.20 & \cellcolor{bad1!25}17.41  &  1914\\
& \textsf{WST-Vamana} & 23.62 & 59.09  &  4556 & 20.56 & 54.04  &  4839 & 2.91 & 9.07 &  1456 & 4.04 & 14.84  &  837\\
& \textsf{iRangeGraph}  & 24.27 & 44.72  &  \cellcolor{bad1!25}69801 & 20.73 & 43.82  &  8455 & 3.33  & 5.30  &  \cellcolor{bad1!25}6731 & 4.66 & \cellcolor{good1!25}7.23  &  \cellcolor{bad1!25}12155\\
& \textsf{UNIFY}  & 30.25 & 36.96  &  17721 & 29.21 & 34.88  &  \cellcolor{bad1!25}24313 & 4.46 & 6.64  &  2308 & 6.36 & 10.48 &  3088\\
\hline
\end{tabular}
\end{table*}

\reviewerthree{Table~\ref{tableindex} reports index size (Size), search memory usage (Mem), and construction time (Time). Blue and red backgrounds indicate the best and worst performance under each filtering method. For \textsf{Milvus}, index size is not directly measurable due to compact storage; memory usage is estimated from total Docker consumption.} 

Compared to \textsf{Faiss-HNSW}, most algorithms exhibit higher index size, memory footprints and construction times. Overall, index size correlates positively with construction time \reviewertwo{(Pearson coefficient$ = 0.71$)}.

\reviewerthree{\ul{\textbf{Index size characteristics.}}} 
Range filtering methods show significant higher index size than Faiss-HNSW, for their theoretical index sizes are \(O(Mn\log(n))\), which is moderate. 
However, in practice, \textsf{SeRF}'s memory usage is even lower, thanks both to its effective search performance with smaller \(M\) values and its tendency to skip many unnecessary ranges during index construction, resulting in a significantly reduced practical memory footprint.

In particular, \textsf{WST-opt} is the most memory-intensive, as it duplicates vector storage across overlapping BST subgraphs. These tradeoffs highlight the balance between filtering capability and storage efficiency, especially as dataset sizes increase.

\ul{\textbf{Memory characteristics.}} 
Filtering ANN methods typically require more memory than naive approaches.In containerized deployments like \textsf{Milvus} standalone, memory usage is strictly limited by Docker's preconfigured allocation regardless of dataset size. Quantization-based methods achieve better memory efficiency through compressed data representations. \textsf{ACORN}, \textsf{Filtered-DiskANN} and \textsf{NHQ-NSW} maintain similar edge counts to HNSW, resulting in comparable index size. 

\reviewerthree{Some methods exhibit disproportionately high memory usage relative to their index sizes due to algorithm designs. For example, \textsf{ACORN} stores a boolean flag per vector for each query, and \textsf{NHQ-KGraph} duplicates the dataset unnecessarily.}



\ul{\textbf{Index construction time.}} 
Construction time is closely correlated with memory usage. An exception is \textsf{iRangeGraph}, which builds subgraphs sequentially (i.e., single-threaded), yielding high-quality indexes at the expense of parallel efficiency. Its reliance on RNG-based graph construction also results in slower indexing compared to methods using Vamana or KGraph. By contrast, \textsf{SeRF} maintains efficient construction times, primarily due to its operation with a small $M$ parameter.




\subsection{Pruning Strategy Evaluation}

\label{sec:exp:prune}


Several Filtering ANN algorithms, such as \textsf{ACORN} and \textsf{Filtered-DiskANN}, modify the standard RNG pruning strategy to improve connectivity under low-selectivity conditions. Similarly, \textsf{NHQ-KGraph} uses a KGraph-style pruning rather than NSW to achieve higher QPS, suggesting that RNG-style pruning may be suboptimal for low-selectivity scenarios. To evaluate this hypothesis, we modified both \textsf{SeRF} and \textsf{ACORN} to incorporate alternative pruning methods and compared their performance using comparisons as a detailed performance indicator. During construction step, ACORN\_kg and SeRF\_kg connect nearest neighbors directly, whereas  ACORN\_rng applies RNG-style pruning method.

\begin{figure}[t]
	\centering
	\begin{subfigure}{\linewidth}
		\centering
		\includegraphics[width=\linewidth]{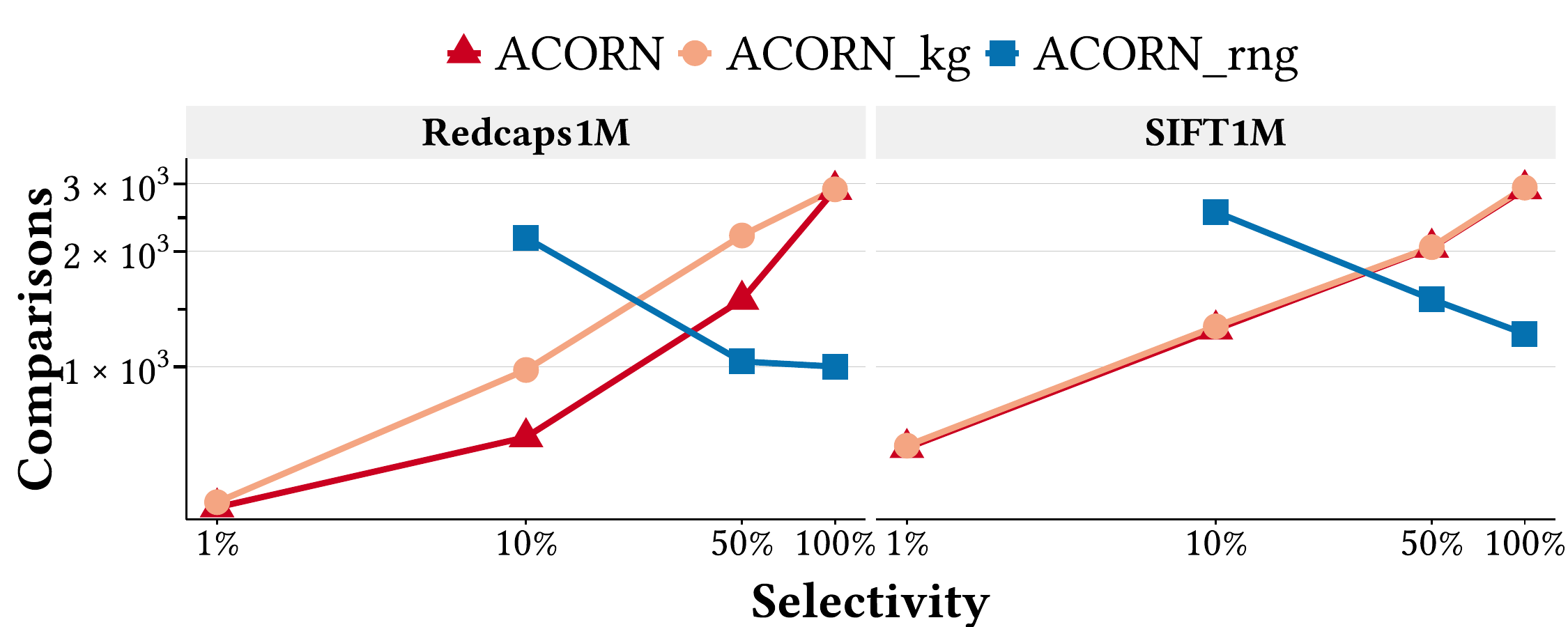}
		\caption{ACORN}
		\label{acornprune}
	\end{subfigure}
    \\
	\centering
	\begin{subfigure}{\linewidth}
		\centering
		\includegraphics[width=\linewidth]{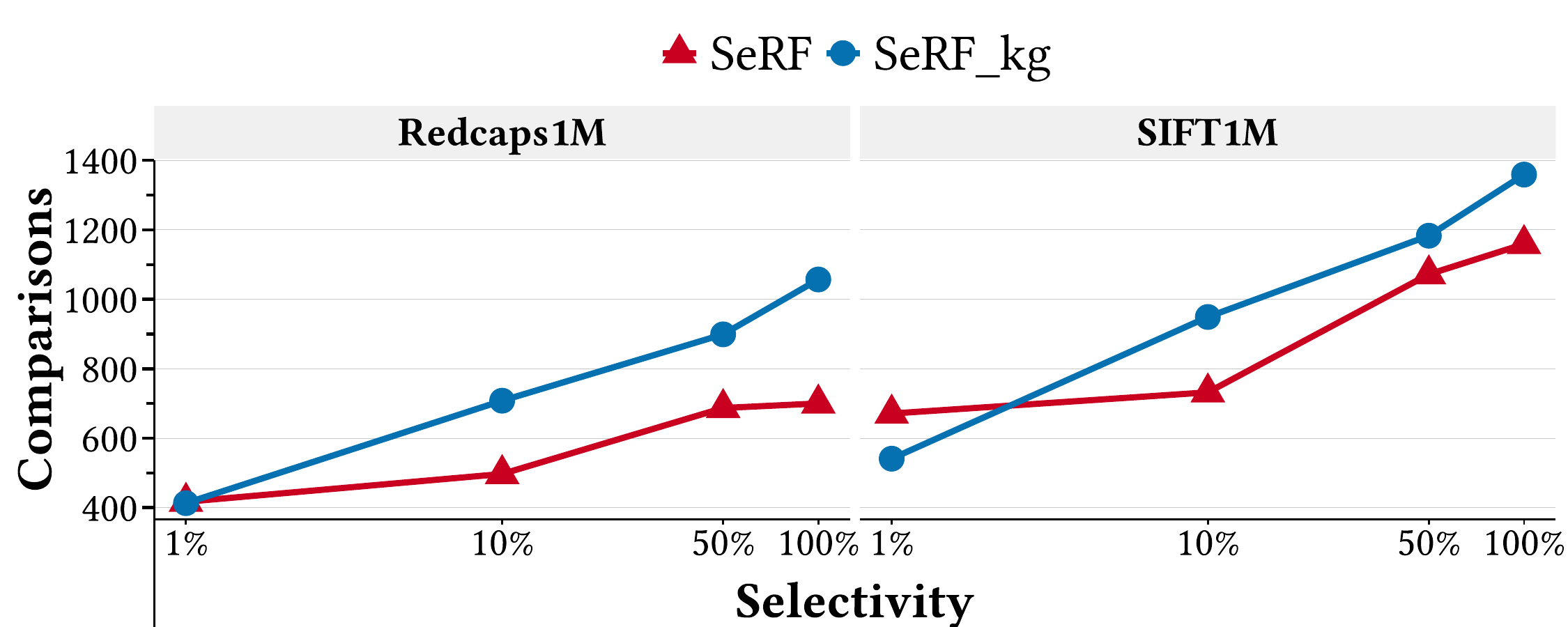}
		\caption{\textsf{SeRF}}
		\label{serfprune}
	\end{subfigure}
        \caption{Comparative performance of pruning methods at 90\% recall\text{@}10 (Fewer \reviewerone{Comparisons} is better).}
	\label{pruneexp}
\end{figure}

Figure \ref{acornprune} shows the performance of different pruning methods across selectivity levels from 1\% to 100\% in \textsf{ACORN} (all methods failed at 0.1\% selectivity). The results indicate that \textsf{ACORN}'s original two-hop pruning performs best for its architecture with low selectivity, while \textsf{ACORN\_kg} achieves similar performance. 
Meanwhile, RNG performs better with above 50\% selectivity, indicating different pruning strategies may suit different selectivity. 

Figure~\ref{serfprune} illustrates the impact of KGraph-style pruning on \textsf{SeRF}. While the original \textsf{SeRF} benefits from efficient edge construction through both RNG-style pruning and segmented edge filtering, \textsf{SeRF\_kg} shows improved performance below 5\% selectivity. However, it still fails to perform effectively at 0.1\%. Overall, RNG-style pruning preserves graph quality at moderate to high selectivity but remains less effective in extremely selective cases.

\subsection{Entry Point Selection Analysis}

\label{sec:exp:entry}


\reviewerthree{Entry point selection is a crucial component in graph-based ANN search. However, simple strategies often rival complex ones in Filtering ANN search.}  For example, \textsf{SeRF} and \textsf{DSG} select entry points from the bottom layer and still match the performance of hierarchical methods like \textsf{UNIFY-joint}. In this analysis, we investigate two factors: (1) the impact of hierarchical structures, and (2) the effect of varying the number of entry points in \reviewerthree{single-layer} graph indexes.

\begin{table}[t]
  \caption{\reviewerone{Comparisons} variation w.r.t. various entry point selection strategies on SIFT.}
  \label{entry}
  \begin{tabular}{c|c|c|c|c|c}
    \hline
    \multirow{2}{*}{\textbf{Algorithm}} & \multirow{2}{*}{\textbf{ep}} & \multicolumn{4}{c}{\textbf{Selectivity}} \\
    \cline{3-6}
    & & \textbf{$1\%$} & \textbf{$10\%$} & \textbf{$50\%$}  & \textbf{$100\%$}\\
    \hline
    \hline
    
    \textbf{UNIFY} & \textbf{1} & \textbf{default} & \textbf{default} & \textbf{default} & \textbf{default}\\
UNIFY-B & 3 & -0.14\% & 0.10\% & 1.09\% & 2.21\% \\
UNIFY-B & 30 & -0.32\% & -0.83\% & -0.91\% & -1.42\% \\
    \hline
    \textbf{SeRF} & \textbf{3} & \textbf{default} & \textbf{default} & \textbf{default} & \textbf{default}\\
SeRF & 30 & -3.34\% & -2.52\% & -2.28\% & -1.34\% \\
SeRF & 300 & -5.53\% & -4.19\% & -3.85\% & -2.84\% \\
    \hline
    \textbf{DSG} & \textbf{3} & \textbf{default} & \textbf{default} & \textbf{default} & \textbf{default}\\
DSG & 30 & -2.78\% & -2.55\% & -2.53\% & -2.18\% \\
DSG & 300 & -4.38\% & -4.26\% & -4.32\% & -4.18\% \\

    \hline
    \textbf{UNIFY-B} & \textbf{3} & \textbf{default} & \textbf{default} & \textbf{default} & \textbf{default}\\
UNIFY-B & 30 & -0.18\% & -0.93\% & -1.98\% & -3.56\% \\
UNIFY-B & 300 & -0.30\% & -1.63\% & -3.33\% & -4.23\% \\
    \hline
\end{tabular}
\end{table}


\reviewerthree{\textsf{UNIFY} represents the optimal solution with a hierarchical structure. To evaluate the impact of hierarchy, we compare it with \textsf{UNIFY-B}, which retains only the bottom layer from \textsf{UNIFY}. Since \textsf{SeRF} and \textsf{DSG} are inherently single-layer indexes, we vary their entry point size (\texttt{ep}) to study the effect of the number of entry points.
Table~\ref{entry} presents the variation in Comparisons when adjusting entry point selection methods relative to each algorithm's default.}


\ul{\textbf{Effect of hierarchical structure.}} 
Comparing \textsf{UNIFY} with \textsf{UNIFY-B} reveals that the hierarchical structure has little impact on low-selectivity queries, but plays a more significant role as selectivity increases. 
We reckon that designing a method to reuse top-layer entry points in segmented edge methods, \textsf{SeRF} and \textsf{DSG}, might also lead to performance gains. 

\ul{\textbf{Effect of entry point size.}} 
Increasing the number of entry points for single layer indexes consistently reduces comparisons across all selectivity levels. \textsf{UNIFY-B}, \textsf{SeRF}, and \textsf{DSG} all perform better when using 30 or 300 entry points rather than just 3. Moreover, increasing the entry point set size improves search performance without recall loss. 
Notably, the performance of bottom-layer entry point selection with 30 or 300 points all surpass the default settings.

\reviewerone{Notably, each method is compared with itself under different settings, so the number of comparisons directly reflects QPS. For example, in SeRF, using 300 entry points yields a 5.3\% increase in QPS compared to the original method (which uses 3 entry points) at 50\% selectivity, with a corresponding 3.85\% decrease in Comparisons. Overall, incorporating more entry points properly is a worthwhile consideration for improving performance.}

\subsection{Edge Filtering Overhead Analysis}

\label{sec:exp:edgesel}

\begin{figure}[t]
    \centering
    \includegraphics[width=\linewidth]{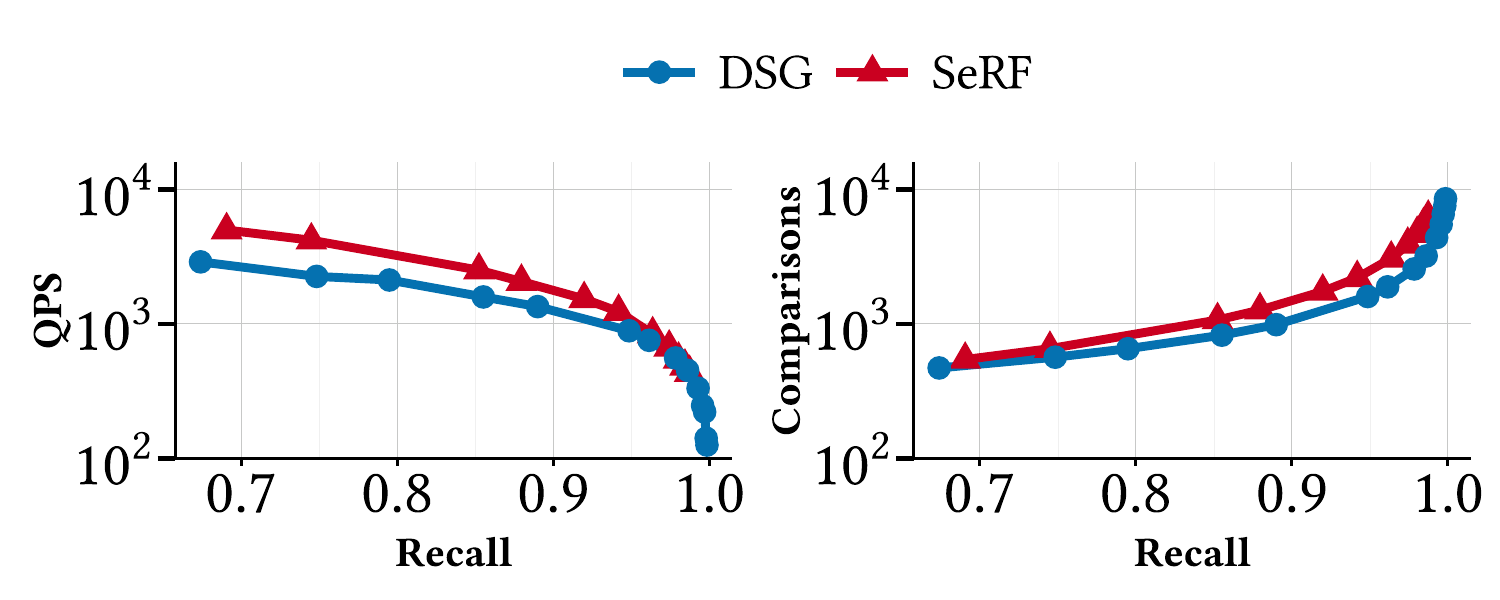}
    \caption{\reviewerone{Recall/QPS and Recall/Comparisons for SeRF and DSG in SIFT at 10\% selectivity.}}
    
    \label{fig:exp_recall_serf_dsg}
\end{figure}


For most graph-based Filtering ANN search methods, comparisons per query show consistent performance, DSG show difference. Compared with \textsf{SeRF}, \reviewerone{\textsf{DSG} performs fewer comparisons per query but achieves lower QPS, as shown in Figure~\ref{fig:exp_recall_serf_dsg}.} 
A potential reason is the additional overhead from filtering during neighbor selection. To validate this, we estimate the percentage of query time spent on neighbor selection.


\begin{figure}[t]
    \centering
    \includegraphics[width=\linewidth]{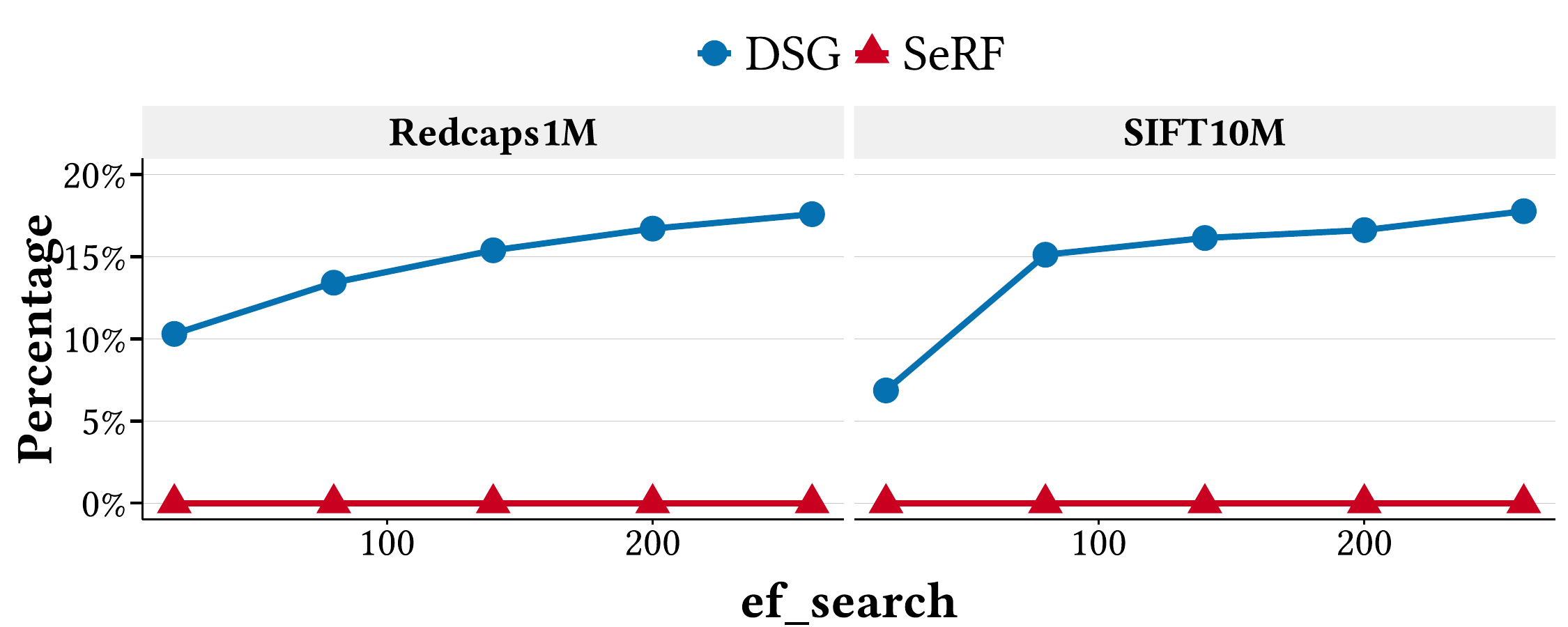}
    \caption{Edge filtering time (\%) in SIFT at 10\% selectivity.}
    
    \label{nns}
\end{figure}

Figure~\ref{nns} demonstrates that \textsf{DSG} spends a significant amount of time finding valid edges, which becomes a key bottleneck. In contrast, \textsf{SeRF} spends almost no time on this step. As {\tt ef\_search} (the size of search candidate set) increases, the time required to filter valid edges also increases. This is mainly because, as the candidate set expands, it needs more time to find matched neighbors.

\section{Lessons Learned} 

\label{sec:lessons}

In this section, we summarize key insights (I) from our study, provide actionable recommendations, and outline promising open problems (O) for future study.

\subsection{Insights}

\textbf{I1. Fine-grained segmented subgraphs are highly effective solutions to range Filtering ANN search.} The most effective approach for range filtered ANN relies on subset indexing, typically implemented as segmented subgraphs (e.g., iRangeGraph, $\beta$-WST, UNIFY). These specialized structures efficiently address the challenges of filtered search by narrowing the search space.

\textbf{I2. Segmented edge-based methods fail at low selectivity due to incomplete search during index construction.}
Methods like SeRF and DSG perform well only when their construction phase successfully connects relevant nodes. In low-selectivity settings, RNG-based techniques often miss such connections due to their monotonic search nature, indicating the need for alternative mechanisms that ensure connectivity.

\textbf{I3. Partitioning is effective for low-selectivity queries.} Milvus adopts partitioning to make low-selectivity queries viable. Similarly, more advanced techniques like stitching and segmented subgraphs further enhance performance in such settings.

\begin{sloppypar}
\textbf{I4. Label filtering algorithms remain underdeveloped.}  
FDiskANN-SVG underperforms compared to NHQ-KGraph at high selectivity, mainly due to the weak quality of the Vamana graph. Conversely, NHQ-KGraph is unreliable at low selectivity. No method performs reliably across all settings, underscoring the need for better label filtering techniques.
\end{sloppypar}


\textbf{I5. Hybrid distance filtering supports only high-selectivity queries.} 
NHQ often fails at low selectivity because its connectivity is not guaranteed across all filtered subsets. Moreover, approaches involving multi-cardinality labels and strict matching are rarely used in practice due to their rigidity.

\textbf{I6. Hierarchical structures offer limited benefits.}
Hierarchical indexing structures show effectiveness only under high selectivity. For example, UNIFY leverages a hierarchical design to sample entry points, but its advantage over the non-hierarchical variant becomes evident only when query selectivity exceeds 50\%.


\textbf{I7. Larger entry point sets improve performance.}
Increasing the number of entry points at the bottom layer consistently improves search performance across various algorithms and selectivity levels, often surpassing hierarchical structures.

\subsection{Tool Selection}

\begin{figure}[t]
  \centering
  \includegraphics[width=\linewidth]{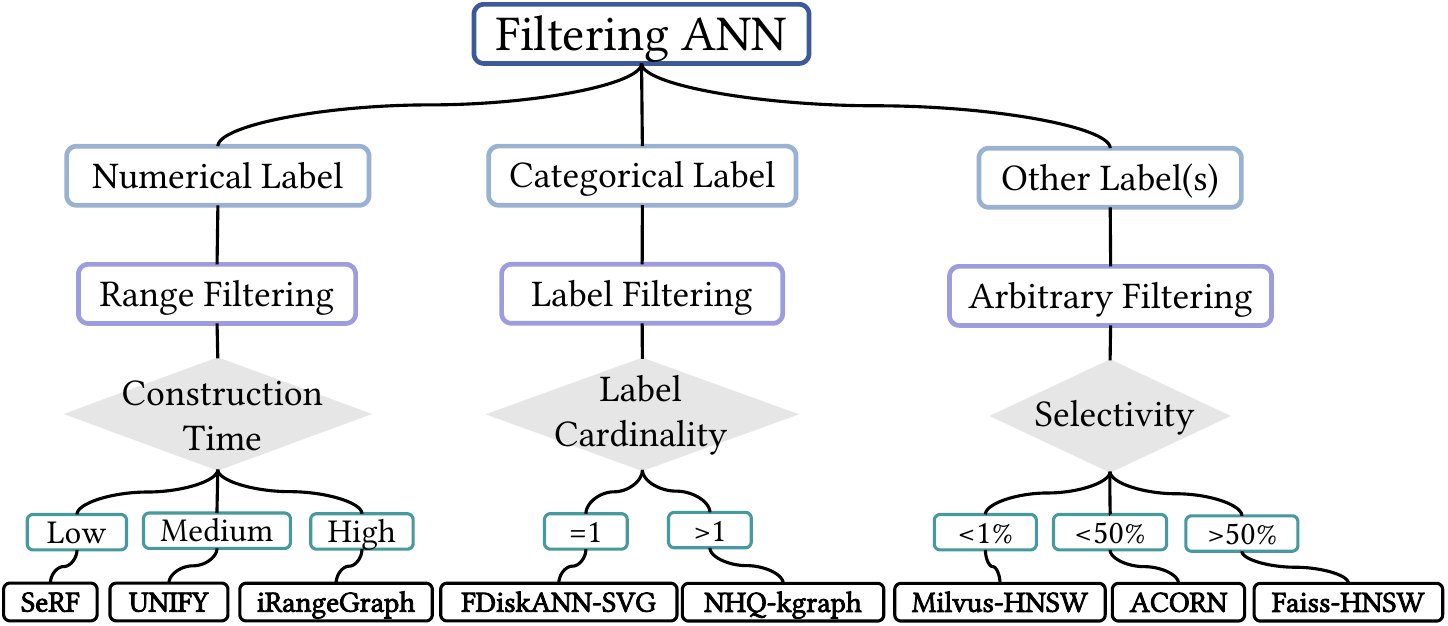}
  \caption{Guidebook for Filtering ANN algorithm usage.}
  \label{guidebook}
\end{figure}

Figure~\ref{guidebook} provides a practical guide for selecting Filtering ANN algorithms based on attribute type, index construction cost, cardinality, and selectivity.

For {\bf numerical attributes}, UNIFY and iRangeGraph are strong options. iRangeGraph typically achieves the best query performance but requires longer index construction time. If construction efficiency or memory is a concern, SeRF offers a faster alternative, though it performs worse at low selectivity. Even though WST-opt achieves comparable performance to iRangeGraph, it is not recommended due to its significantly higher memory consumption. In contrast, while iRangeGraph has slower index construction, mainly because of its low parallelization, which is relatively easy to optimize.

For {\bf label filtering}, FDiskANN-SVG is recommended when the query label cardinality is one. When cardinality exceeds one, NHQ-KGraph offers better performance.

In {\bf arbitrary filtering} scenarios involving complex or mixed conditions, ACORN is the most flexible solution. However, for queries with high selectivity (>50\%), traditional methods like Faiss-HNSW are often more efficient. For very low selectivity (<1\%), \textsf{Milvus-HNSW} is preferred because of its stable and efficient scan strategy. 





\subsection{Open Problems}

\textbf{O1. Arbitrary Filtering ANN search remains an open challenge.}
ACORN is currently the only method designed for arbitrary filtering, yet it struggles with robustness across varying selectivity levels. The core difficulty lies in supporting arbitrary subset queries, which makes subgraph-based approaches (e.g., segmented or stitched methods) difficult to apply effectively. A promising direction is to explore IVF-based techniques, whose inherent partitioning structures may better support flexible and efficient subset search.

\textbf{O2. Achieving optimal performance requires careful hyperparameter tuning.}
Filtering ANN search performance is highly sensitive to hyperparameters in both index construction and query execution. Optimal settings vary based on factors such as (1) dataset size, (2) vector distribution, (3) query selectivity, (4) size of $K$ to be retrieved, and (5) intrinsic dimensionality for quantized methods. This makes universal tuning impractical. Future work should explore adaptive tuning strategies or develop robust algorithms that perform well without extensive manual calibration.

\textbf{O3. Integration with vector databases remains limited.}
Most Filtering ANN methods rely on specialized index architectures tailored for specific filtering tasks (e.g., range or label filtering), creating barriers to integration with general-purpose vector databases. There is a clear need for hybrid indexing designs that retain compatibility with traditional ANN structures while supporting efficient filtering, enabling seamless adoption in real-world database systems.

\textbf{O4. Dynamic Filtering ANN index is still underexplored.}
Among existing methods, only DSG supports dynamic index updates. Other range filtering approaches—such as SeRF, $\beta$-WST, iRangeGraph, and UNIFY—depend on attribute-sorted vector orders and do not support incremental insertions or deletions. Overcoming this limitation remains a fundamental challenge for making Filtering ANN algorithms suitable for dynamic environments.

\reviewerthree{\textbf{O5. The attribute distribution analysis lacks.} 
All generated attributes are assumed to follow a uniform distribution, a setting adopted by most existing methods. However, this assumption may not reflect real-world scenarios. 
Some works such as \textsf{ACORN}~\cite{patel2024acorn} and \textsf{AIRSHIP}~\cite{zhao2022constrained} acknowledge this issue, 
but a comprehensive analysis of the interaction between vectors and attributes, as well as the performance of different methods under varying vector-attribute distributions, is still lacking.
}

\section{Related Work}

In addition to the algorithms and systems discussed in this paper, several other notable methods have been proposed. However, they are excluded from our experiments due to lower efficiency reported in prior studies or lack of open-source implementation.

\textbf{Vector database systems.}
\textit{AnalyticDB-V (ADBV)}~\cite{wei2020analyticdb} is a vector database system that supports vector search under arbitrary restrictions. ADBV employs a \textit{Voronoi graph (VoG)} as an IVF structure to reduce scan space and combines it with product quantization (PQ) for memory efficiency. A built-in cost estimator selects the most efficient strategy from among brute-force scan, PQ pre-filtering, VoGPQ pre-filtering, and VoGPQ post-filtering. 



\textit{VBASE}~\cite{zhang2023vbase} is a vector database developed by Microsoft, extending PostgreSQL~\cite{postgresql} with SQL-based vector search. It introduces a termination signal mechanism called \textit{Relaxed Monotonicity}, which halts the search once retrieved candidates begin to diverge from the query vector. For filtering queries, VBASE softens the constraints to expand the search space, ensuring sufficient matched results. 

Several other vector database systems are not included in our study, such as Vearch~\cite{vearch, li2018design}, PASE~\cite{yang2020pase}, Weaviate~\cite{weaviate}, Pinecone~\cite{pinecone}, and Qdrant~\cite{qdrant}. Among all vector database systems, \reviewertwo{Milvus is the most competitive method for its high efficiency, versatility~\cite{pan2024survey} and widespread LLM applications~\cite{edge2024local}.} Hence, we pick Milvus as the representative vector database system solution.




\textbf{Filtering ANN algorithms.}
\textit{RII}~\cite{matsui2018reconfigurable} introduced the concept of subset search, a precursor to filtered search. Built on IVFPQ, RII uses two scanning strategies based on a threshold: if the subset size is small, it scans the subset directly to get results; otherwise, it employs id-IVF, an index structure that maps IDs to clusters and skips clusters that do not contain target IDs. 
\reviewertwo{RII is omitted due to its naive filtering strategy, and its performance is almost identical to Faiss-IVFPQ according to our preliminary test.}




\textit{AIRSHIP~\cite{zhao2022constrained}} is a label Filtering ANN method built on HNSW. It assumes that attribute distributions are biased and clustered, and samples vectors with diverse labels during index construction to ensure entry point coverage for all label types. 
During query execution, it performs both label-constrained and unconstrained searches using two separate candidate lists,
making AIRSHIP effective under skewed attribute distributions. \reviewertwo{AIRSHIP was excluded from our experiments due to the unavailability of its open-source code}.

\section{Conclusion}

In this paper, we present a comprehensive taxonomy and empirical study of filtering approximate nearest neighbor (ANN) algorithms, analyzing both algorithmic principles and implementation details. We categorize Filtering ANN methods based on the type of attributes, i.e., numerical (range filtering) and categorical (label filtering), and highlight the growing trend toward arbitrary filtering techniques for supporting arbitrary constraints. From a design standpoint, we classify filtering strategies into pre-filtering, post-filtering, and joint-filtering, each representing distinct trade-offs. For range filtering, we further examine implementation-level design choices that influence performance.

Our extensive experiments across diverse query scenarios show that segmented subgraph indexes consistently achieve superior performance in range filtering tasks. In contrast, label filtering remains unstable due to the lack of high-quality graph indexes. We identify two key factors limiting edge-based indexes: (1) pruning strategies and (2) entry point selection. Our study offers practical guidance for selecting suitable Filtering ANN methods and outlines open challenges that point toward promising directions for future research.

\nocite{latex}
\bibliographystyle{ACM-Reference-Format}
\bibliography{sample-base}

\appendix

\end{document}